\documentclass[%
 reprint,
 amsmath,amssymb,
 aps,
prb,
]{revtex4-2}

\usepackage{graphicx}
\usepackage{dcolumn}
\usepackage{bm}
\usepackage{amsmath}

\usepackage{tabularx}

\usepackage{hyperref}
\hypersetup{
    colorlinks=true,
    linkcolor=blue,
    urlcolor=blue,
    citecolor=red
}

\usepackage[deletedmarkup=xout]{changes}

\begin{document}

\preprint{APS/123-QED}

\title{Avalanche Dynamics and the Effect of Straining in Dislocation Systems with Quenched Disorder}

\author{D\'enes Berta}
\author{Barna Mendei}
\author{P\'eter Dus\'an Isp\'anovity}
\affiliation{ELTE E\"otv\"os Lor\'and University, Department of Materials Physics, P\'azm\'any P\'eter s\'et\'any 1/A, 1117 Budapest, Hungary}


\begin{abstract}
The plastic deformation of crystalline and other heterogeneous materials often manifests in stochastic intermittent events indicating the criticality of plastic behavior. Previous studies demonstrated that the presence of short-ranged quenched disorder modifies this behavior disrupting long-range static and dynamic correlations consequently localizing dislocation avalanches. However, these observations were mostly confined to relaxed materials devoid of deformation history. In this work our focus is on how straining affects static and dynamic correlations, avalanche dynamics and local yield stresses. We demonstrate that the interplay between severe straining and confining quenched disorder induces critical behavior characterized by dislocation avalanches distinct from those at lower stresses. Namely, near the flow stress many avalanches, even if triggered locally, evolve into events affecting a larger region by exciting small clusters of dislocations all around the sample. This type of avalanches differ from the ones at low strains where plastic events typically consist of one compact cluster of dislocations which is either local or it is already quite extended at the onset of the avalanche. Furthermore, we examine the impact of avalanches on local yield stresses. It is shown in detail in this work that while some statistical features of the local yield thresholds are robust to straining, others are significantly affected by the deformation history.
\end{abstract}

\maketitle

\section{Introduction}

Previous experimental results revealed that the plastic behavior of micron and sub-micron scale crystalline specimens fundamentally differ from that of their bulk counterparts. In this regime size-related hardening is observable \cite{fleck1994strain, uchic2004sample, uchic2009plasticity}, the deformation manifests as stochastic sequence of strain bursts \cite{dimiduk2006scale, csikor2007dislocation} that are often also accompanied by acoustic emission 
\cite{weiss1997acoustic, weiss2003three,weiss2015mild,ispanovity2022dislocation}. This stochasticity results in unpredictable plasticity and a zig-zag pattern in the stress-strain curves in contrast with the smooth behavior of bulk samples. In crystalline materials these fluctuations are results of sudden rearrangements of dislocations known as dislocation avalanches. Similar behavior characterizes other heterogeneous materials such as amorphous materials or foams in which the fluctuations are related to shear transformation zones \cite{spaepen1977microscopic, falk1998dynamics} and T1 events \cite{kabla2003local, dennin2004statistics, tainio2021predicting}, respectively.

Despite the stochastic and seemingly unpredictable nature of plasticity of heterogeneous materials, several attempts have been made to establish a direct connection between the features of the microstructure and the emergent plastic response \cite{richard2020predicting}. One of the most powerful predictor of the plastic behavior is the so-called \emph{local yield stress}, that is, the critical stress at which the material yields during local probing. This descriptor was first showed to be a powerful tool of predicting the loci of plastic events in model amorphous solids \cite{patinet2016connecting} and then its applicability was demonstrated in crystalline materials modeled by discrete dislocation systems in both 2D \cite{berta2023dynamic} and 3D \cite{berta2025identifying} as well. Numerical studies on metallic glasses revealed that the statistics of local yield stresses strongly depend on the preparation protocol \cite{barbot2018local}. In the case of crystalline materials it was shown that global plasticity is related to the local yield stresses through the weakest-link principle, however, this weakest-link picture is fundamentally different depending on whether long- or short-range interactions dominate the dislocation dynamics \cite{berta2023dynamic}. Namely, in systems dominated by long-range elastic dislocation-dislocation interactions plastic events are spatially extended which leads to a weakest-link behavior of moderate consistency. If, however, a significant extent of short-ranged quenched disorder (e.g.,\ point defects) is introduced, the plastic events get localized and a more rigorous traditional weakest-link picture is realized. This is related to the concept of the \emph{wild to mild} transition which explains the empirical observation that depending on the interaction involved a material my exhibit very intense or very mitigated acoustic emission fluctuations during deformation \cite{weiss2015mild}.

As it was discussed above, the local yield stress is a very important indicator of the plasticity of materials and it connects the plastic behavior with structural properties through the weakest-link principle. This makes local yield stresses very important ingredients of mesoscale models because they can be utilized to introduce local strength fluctuations characteristic to heterogeneous materials \cite{morris2011size, liu2018elastic}. While it was shown for amorphous solids that the local yield thresholds correlate with plastic activity even after several plastic rearrangements it is still unknown how local yield stresses evolve if materials are exposed to significant deformation. Its importance from the mesoscopic modeling point of view comes from the need to understand how the local strengths have to be updated as the system evolves and deformation occurs and how their statistics is influenced by the deformation history. In this work we investigate these questions as well as the effect of deformation history in the framework of 2D discrete dislocation dynamics (DDD). One of the other foci of this work is how static (structural) and dynamic correlations and avalanche dynamics are affected by straining (nearly up to the phase transition to sustained plastic flow) as well as by subsequent unloading of the sample.

The paper is structured as follows. After a summary of the model applied, the static correlations are investigated, then the focus is moved to the dynamic correlations. This is followed by an analysis of the plastic events carried out on the level of individual dislocations, and three distinct regimes of avalanche behavior are identified. Finally, local yield stresses are studied and the paper is concluded with a discussion and an outlook.

\begin{figure}[ht!]
    \centering
    \includegraphics[width=\columnwidth]{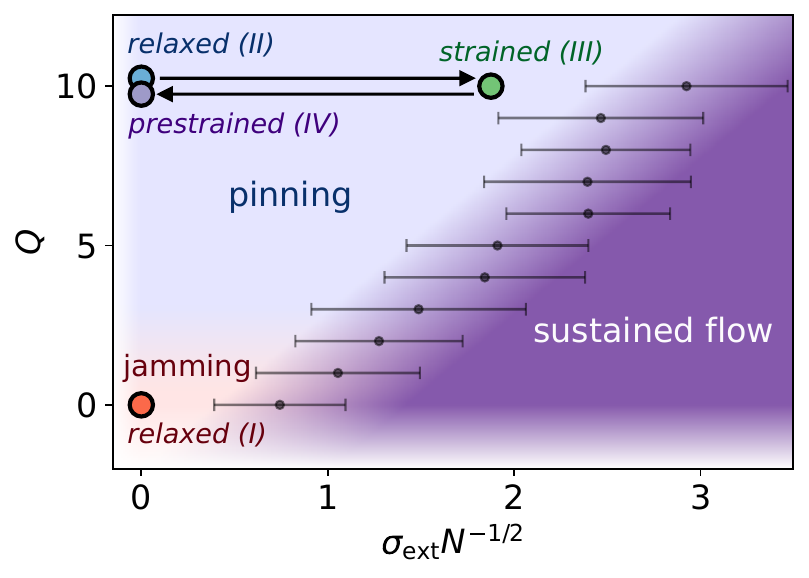}
    \caption{The phase diagram of 2D discrete dislocation systems with quenched disorder. $\sigma_\mathrm{ext}$ and $N$ are the external stress and the number of dislocations, respectively. $Q$ is the ratio of the number of point defects and dislocations characterizing the extent of quenched disorder in the system. The markers indicate the four cases that are investigated in this paper. The relaxed $Q=0$ system (condition I) is governed by long-range dislocation-dislocation interactions and the case of dislocation jamming is realized. The $Q=10$ systems are dominated by the short-range point defect - dislocation interactions, that is, it lies in the pinning regime. Besides the relaxed $Q=10$ configurations (condition II) two other types of $Q=10$ systems were studied. The configurations were first strained close to the phase transition to sustained flow (condition III). Then, the configurations are unloaded to zero external stress (condition IV). The errorbars indicate the normalized stress at which the configuration enters the state of sustained plastic flow.}
    \label{fig:phase-space}
\end{figure}

\section{Simulation methods}

In order to study the issues outlined in the introduction a 2D discrete dislocation dynamics (DDD) model is applied. This model, despite its simplicity, has proven to capture the main features of the criticality of crystalline plasticity since it was shown to be statistically consistent with experiments in terms of strain burst distribution, time distribution of dislocation avalanches, etc. \cite{csikor2007dislocation, ispanovity2022dislocation}. Additionally, the relatively simple setup (to be discussed in the next paragraph) allows us to study the large-strain regime (even close to the phase of sustained plastic flow) which is immensely more challenging with more complex 3D DDD models or lower scale molecular dynamics (MD) approaches that are typically limited to small strains and/or system sizes.

In the model a square-shaped ($L\times L$ sized) simulation cell is considered that contains $N=1024$ edge dislocations with line directions perpendicular to the $xy$ plane of the simulation cell. Half of the dislocations have a Burgers vector of $\bm b=b\bm e_x$ and the other half have $\bm b=-b\bm e_x$  where $b$ is the magnitude of the Burgers vectors and $\bm e_x$ is the unit vector in direction $x$. The motion of the dislocations is restricted to slip motion in the direction $x$ and they interact via a shear stress field of 
\begin{equation}
    \tau_\mathrm{d}(\bm r) = \frac{\mu b}{2\pi(1-\nu)}\frac{x\left(x^2-y^2\right)}{\left(x^2+y^2\right)^2}
    \label{eq:stress_field}
\end{equation}
where $\bm r=(x,y)$ is the relative position with respect to a dislocation and $\mu$ and $\nu$ are the shear modulus and the Poisson's ratio, respectively. We note that in the paper stresses are normalized by stress units $\tau_0=\frac{\mu b \sqrt{\rho}}{2\pi(1-\nu)}$ where $\rho=N/L^2$ is the dislocation density. Besides dislocations, the configurations may contain point defects introducing short-ranged quenched disorder to the system. The point defects are immobile but they interact with the dislocations via their shear stress fields of  
\begin{equation}
    \tau_\mathrm{v}(\bm r) = -2Axy\frac{\frac{1-\mathrm{exp}\left[-K^2r^2\right]}{r^2}-K^2\mathrm{exp}\left[-K^2r^2\right]}{r^2}
\end{equation}
where $\bm r=(x,y)$ is the relative position with respect to the point defect, $A$ and $K$ are constants characterizing the strength and the range of the stress field of the point defect. Based on Ref. \cite{ovaska2015quenched} these parameters were chosen as $A=0.0016~\tau_0L^2/\sqrt{N}$ and $K=103.125~\sqrt{N}/L$. The extent of quenched disorder is quantified by the ratio $Q=N_\mathrm{p}/N$ of the number of point defects $N_\mathrm{p}$ and the number of dislocations $N$. In our simulations $Q$ is either 0 (no point defects) or 10 (10240 point defects in each configuration). While several novel methods have been developed to handle boundary conditions \cite{pan2017generalized, berta2020efficient, shima2022nonsingular}, surface effects are outside of the scope of this work, therefore, periodic boundary conditions (PBC) are applied for simplicity. PBC was implemented according to the procedure described in Ref. \cite{peterffy2020efficient}. The dynamics of dislocations is governed by the dislocation-dislocation interactions, point defect - dislocation interactions and potentially homogeneous external stress acting on the dislocations. The emerging stiff system of differential equations is solved with an efficient implicit numerical scheme described in detail in Ref. \cite{peterffy2020efficient} that provides a solution with practically no error beyond numerical precision.

\begin{figure}[ht!]
    \centering
    \includegraphics[width=\columnwidth]{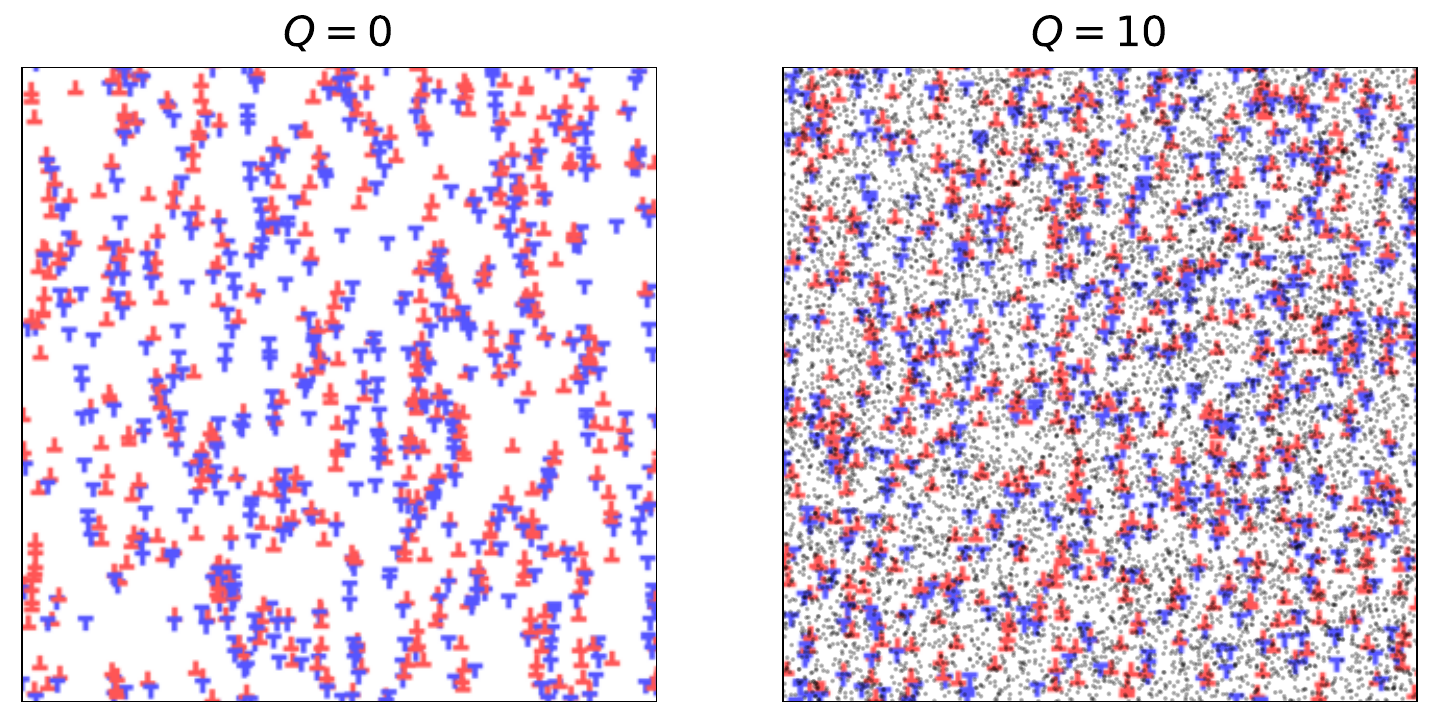}
    \caption{Two representative configurations of relaxed discrete dislocation systems for cases $Q=0$ and $Q=10$ where $Q$ is the ratio of the number of point defects and dislocations. `T' shaped markers indicate the positions of edge dislocations red and blue ones corresponding to dislocations with Burgers vectors pointing to the positive and the negative $x$ direction, respectively. The small gray markers denote the position of the point defects.}
    \label{fig:model_sketch}
\end{figure}

The configurations are prepared in the following way. The dislocations (and point defects) are positioned randomly according to a 2D uniform distribution. Then, the configuration is let to relax and find a metastable configuration while no external stress is applied. This relaxation may be followed by subsequent loading as described later. In our work we studied the ensembles of 100 configurations of each type (see examples of relaxed configurations in Fig.~\ref{fig:model_sketch}). The states at which the behavior of the configurations are investigated are summarized in Fig.~\ref{fig:phase-space}. In total four different conditions are studied (which are referred to as conditions I, II, III and IV later in the paper).
\begin{itemize}
    \item[(I)] Configurations without point defects (${Q=0}$) relaxed at zero external stress.
    \item[(II)] Configurations with point defects (${Q=10}$) relaxed at zero external stress.
    \item[(III)] Configurations with point defects (${Q=10}$) that are strained close to the state of sustained plastic flow by applying a spatially homogeneous external stress $\sigma_\mathrm{ext}=1.875\tau_0\sqrt{N}$. The strained state is achieved by instantaneously increasing the external stress from zero to the desired value and then letting the system to find a new equilibrium state.
    \item[(IV)] Prestrained configurations with point defects (${Q=10}$) that are obtained from condition III by  instantaneously unloading the systems (that is, decreasing the external stress to zero) and letting them to relax. These configurations only differ from condition II in their prestraining history.
\end{itemize}
The external stress value $\sigma_\mathrm{ext}$ was chosen based on preliminary investigations which were conducted to find the flow stress $\sigma_\mathrm{flow}$ at which the system enters the regime of sustained plastic flow. To this end, simulations of 100-dislocation systems were performed. $Q$ was varied as $Q=0,1,2,...,9,10$ and an ensemble of 10 individual simulations were considered for each value of $Q$. After a relaxation step (same as it is described above in the case of $N=1024$ configurations) the configurations were loaded slowly with constant stress rate and homogeneous external stress. The value of $\sigma_\mathrm{flow}$ was determined by tracking the mean dislocation velocity because when entering the sustained flow regime the mean dislocation velocity does not decay to zero any longer but takes non-zero values with periodic fluctuations due to the PBC. Based on the scale-invariant nature of our discrete dislocation model the obtained flow stress is universal in the sense that $\sigma_\mathrm{flow}/\sqrt{N}$ is the same  (for a given $Q$) for systems of different size unless $N$ is very small\cite{tsekenis2011dislocations,szabo2015plastic}. The mean values of the rescaled flow stresses are indicated by the gray markers in Fig.~1 and the error-bars denote the standard deviation corresponding to configuration-to-configuration variation. The value of $\sigma_\mathrm{ext}$ in condition III was chosen based on this prior analysis so the system is strained close to the phase transition but it is still below the regime of sustained plastic flow.

   The dynamical behavior and local yield stresses of these four conditions listed above were studied by examining the avalanches triggered by quasi-static loading with a spatially homogeneous external stress. An avalanche is considered to have started when the mean dislocation velocity exceeds a predefined threshold $v_\mathrm{thr}=10^{-4}v_0$. $v_0=\frac{\mu b^2\sqrt{\rho}}{2\pi(1-\nu)B}$ is the unit of velocities in our simulations where $\mu$ is the shear modulus, $b$ is the length of the Burgers vector, $\rho$ is the dislocation density, $\nu$ is the Poisson ratio and $B$ the dislocation drag coefficient characterizing dislocation mobility. When the mean velocity drops again below this threshold, the avalanches are considered to have ceased (see Fig. \ref{fig:2_velocities}). For more details of the loading and avalanche detection protocol during local yield stress measurements see the corresponding section on local yield stresses and Ref. \cite{berta2020efficient}.

\section{Results}

\subsection{Static correlations}

\begin{figure}[ht!]
    \centering
    \includegraphics[width=\columnwidth]{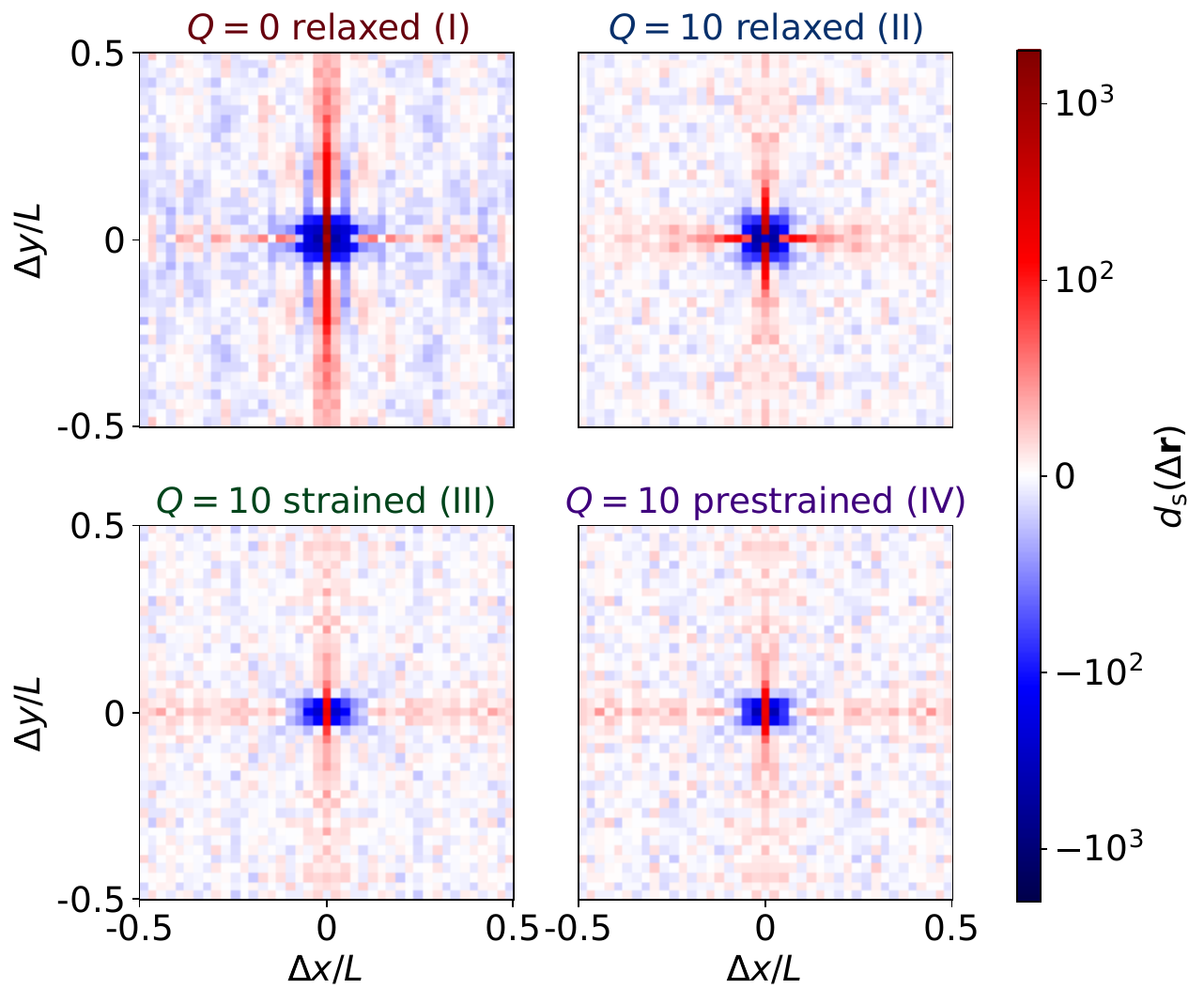}
    \caption{The two-point correlation $d_\mathrm{s}$ of same-signed dislocations (for the definition see Eq. (\ref{eq:static})). In the horizontal and vertical directions (in the latter, in particular) the dislocation configurations are strongly correlated especially in systems without quenched disorder ($Q=0$).}
    \label{fig:correlation_maps}
\end{figure}

\begin{figure}[ht!]
    \centering
    \includegraphics[width=\columnwidth]{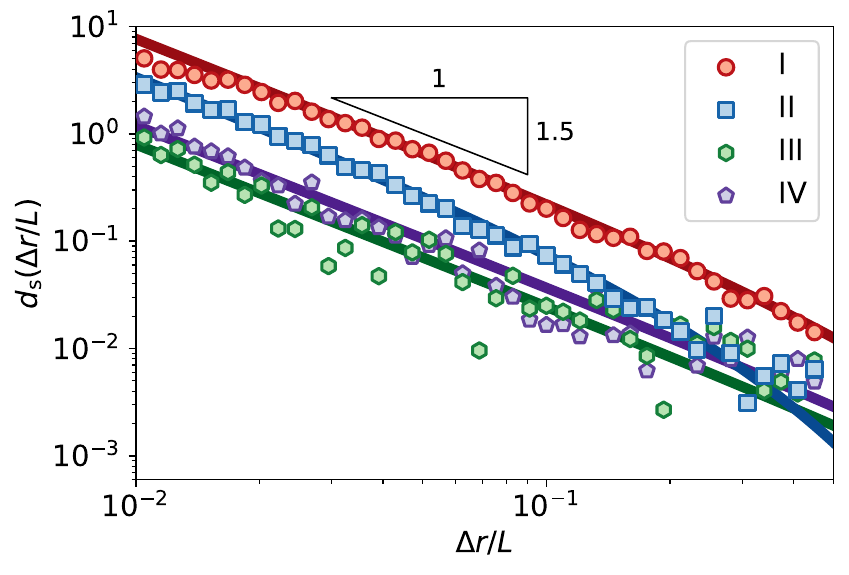}
    \caption{The two-point correlation $d_\mathrm{s}$ of same-signed dislocations in the vertical direction [for the definition see Eq. (\ref{eq:static})]. $\Delta r/L$ is the distance of dislocations normalized with the simulation box size. The correlation decays with an exponent of 1.5. In condition II the correlation has an exponential cutoff characterized by the static correlation length $\xi_\mathrm{s}\approx L/4$. In the other three conditions the data suggest that the cutoff is much larger ($\xi_\mathrm{s}\gtrapprox L$) even though it cannot be determined precisely for conditions III and IV. The solid lines indicate fitted functions  [see Eq.~(\ref{eq:cutoff})].}
    \label{fig:correlation_vertical}
\end{figure}

The static correlations of the dislocation configurations are studied by computing the two-point correlation function
\begin{equation}
    d_\mathrm{s}(\Delta\bm r) = \frac{\rho_\mathrm{s}^{(2)}(\Delta\bm r)}{\left\langle\rho_\mathrm{s}^{(2)}\right\rangle}-1
    \label{eq:static}
\end{equation}
where $\Delta\bm r$ is the relative position of two dislocations of the same sign and $\rho_\mathrm{s}^{(2)}$ is the two-point density of same-signed dislocations. $d>0$ means that the relative position is more frequent than in a completely random configuration and $d<0$ corresponds to a relative position which is less likely. The correlation maps are shown in Fig. \ref{fig:correlation_maps}. It is obvious that horizontally and especially vertically there is a strong correlation of dislocations (that is, a dislocation favors a position below or above another one of the same sign). We note that the correlation of opposite-signed dislocations was previously shown to be shorter-range \cite{zaiser2001statistical, ispanovity2008evolution}, therefore, we only focus on the same-signed correlations in this work. In order to characterize the range of static correlations, the spatial dependence of the correlation function was computed in the vertical direction within a narrow cone of central angle of $\varphi=2\cdot\mathrm{tan}^{-1}(1/10)\approx 11.4^\circ$. The correlation functions in the vertical direction obeys 
\begin{equation}
    d_\mathrm{s}(\Delta r) \propto (\Delta r)^{-\alpha}\mathrm{exp}\left[{-\frac{\Delta r}{\xi_\mathrm{s}}}\right]
    \label{eq:cutoff}
\end{equation}
where $\alpha\approx1.5$ (as derived in Ref. \cite{groma2006debye} using a variational approach) for all four distinct conditions investigated but the static correlation length $\xi_\mathrm{s}$ is affected by the extent $Q$ of quenched disorder and the straining history (see Fig. \ref{fig:correlation_vertical}). The correlations are the strongest in $Q=0$ systems (condition I) and the data suggests that $d_\mathrm{s}(\Delta r)$ has a large cutoff length ($\xi_\mathrm{s}\gtrapprox L$). However, long-range correlations are disrupted by the quenched disorder leading to a static correlation length of $\xi\approx L/4$ in condition II. One could expect that these correlations get stronger and the static correlation length diverges during straining as the system approaches the phase-transition to the sustained flow state. The prefactor of the dislocation-dislocation correlation, however, gets smaller as the system gets severely strained (condition III) and it remains small as the system is unloaded (condition IV). In the case of conditions III and IV the exponential cutoff cannot be exactly determined due to the scatter, however, the cutoff in $d_\mathrm{s}$ is clearly larger than in condition II consistently with the expected divergence at the critical point.

\subsection{Dynamic correlations}

In the previous section the static (configurational) correlations were studied. In this section we address the issue whether and how the deformation state and deformation history affects the dynamics of the system. To this end we define the two-point velocity correlation $\langle v_iv_j\rangle$ as,
\begin{equation}
    \langle v_iv_j\rangle( \Delta \bm r) =  \left\langle\sum_{i=1}^{N_d}\sum_{\substack{j=1, \\ j\neq i}}^{N_d}{\delta_\mathrm{D}(\Delta \bm r - \bm r_i + \bm r_j)v_iv_j}  \right\rangle
\end{equation}
where $N_d$, $\bm r$ and $v$ are the number of dislocations, their position vector and magnitude of velocity, respectively. $\delta_\mathrm{D}$ denotes the Dirac delta generalized function. In a similar fashion the two-point displacement correlation $\langle \Delta x_i\Delta x_j\rangle$ is defined as
\begin{equation}
    \langle \Delta x_i\Delta x_j\rangle(\Delta \bm r) = \left\langle\sum_{i=1}^{N_d}\sum_{\substack{j=1,\\ j\neq i}}^{N_d}{\delta_\mathrm{D}(\Delta \bm r - \bm r_i + 
\bm r_j)\Delta x_i\Delta x_j}\right\rangle
\end{equation}
where $\Delta x$ is the displacement of an individual dislocation.

\begin{figure}[ht!]
    \centering
    \includegraphics[width=\columnwidth]{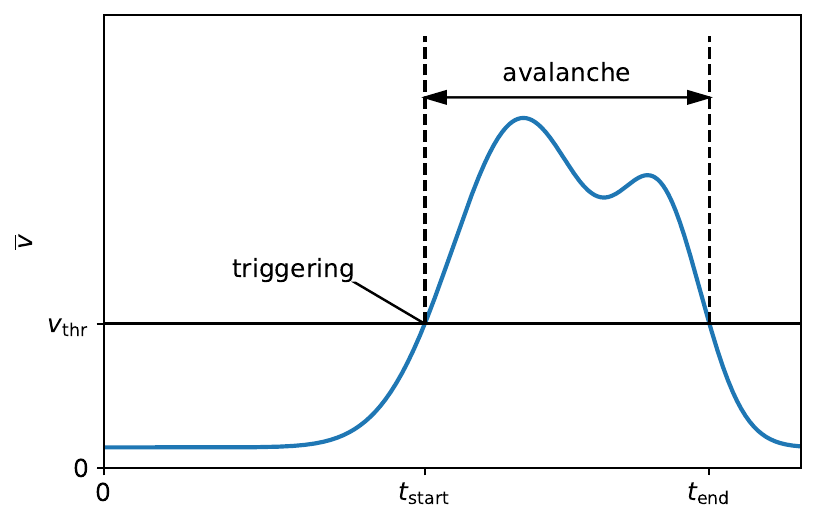}
    \caption{The schematic illustration of how the correlations $\langle v_iv_j\rangle$ or $\langle\Delta x_i\Delta x_j\rangle$ are evaluated. $\langle v_iv_j\rangle$ is computed at time $t_\mathrm{start}$, that is, at the onset of the avalanche, when the mean dislocation velocity $\overline{v}$ exceeds a predefined threshold $v_\mathrm{thr}$. $\langle\Delta x_i\Delta x_j\rangle$ is computed for the duration of the avalanche. This is the time interval starting at $t_\mathrm{start}$ and ending at time $t_\mathrm{end}$ when $\overline{v}$ drops below $v_\mathrm{thr}$.}
    \label{fig:2_velocities}
\end{figure}

$\langle v_iv_j\rangle$ is computed for instantaneous velocities of individual dislocations and $\langle \Delta x_i\Delta x_j\rangle$ is evaluated for their displacements during given time intervals. Dislocation positions are evaluated at the moment at which the instantaneous velocities are considered or at the beginning of the studied interval in the case of displacement correlations. In the following $\langle v_iv_j\rangle$ is evaluated at the triggering of the first avalanche (that is when the mean dislocation velocity exceeds the a predefined threshold $v_\mathrm{thr}$) and $\langle \Delta x_i\Delta x_j\rangle$ is evaluated during the avalanche (i.e., for the interval starting at the triggering and ending when the mean dislocation velocity drops below $v_\mathrm{thr}$, see Fig. \ref{fig:2_velocities}). In this section the long-range asymptotic behavior of $\langle v_iv_j\rangle$ and $\langle \Delta x_i\Delta x_j\rangle$ is under inspection. We note that the instantaneous velocity and displacement correlations can be also interpreted as correlations of plastic strain rate and correlations of accumulated plastic strain contributions of individual dislocations, respectively.

\begin{figure}[ht!]
    \centering
    \includegraphics[width=\columnwidth]{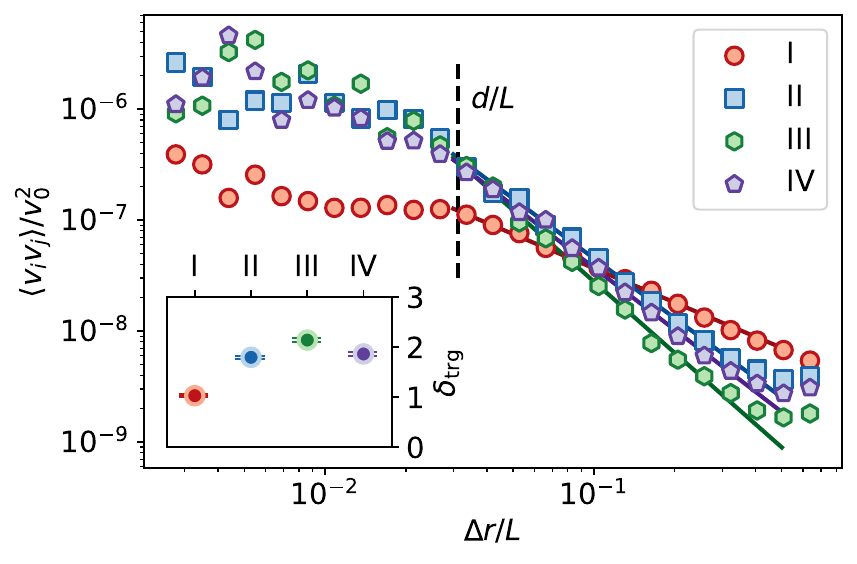}
    \caption{The velocity correlation $\langle v_iv_j\rangle$ normalized with $v_0^2$ at the triggering of the first avalanche. For the definition of $v_0$ see the main text. The inset shows the exponents $\delta_\mathrm{trg}$ defined in Eq. \ref{eq:vv_exp}. The dashed line indicates $d/L$ which denotes the average dislocation spacing normalized with the linear box size $L$.}
    \label{fig:triggering}
\end{figure}

At the onset of the first avalanche (during triggering) the asymptotic behavior is
\begin{equation}
    \langle v_iv_j \rangle(\Delta \bm r) \propto \Delta r ^{-\delta_\mathrm{trg}}
    \label{eq:vv_exp}
\end{equation}
where $\Delta r=|\Delta \bm r|$ is the distance of two dislocations. Figure \ref{fig:triggering} shows that in the critical $Q=0$ case (condition I) the triggered dislocation clusters are very extended already upon the onset of the avalanche with a low value of $\delta_\mathrm{trg}\approx1.0$ (see inset). In the presence of point defects (condition II), however, the decay of $\langle v_iv_j\rangle$ is much faster with $\delta_\mathrm{tgr}\approx1.8$. Perhaps contra-intuitively during straining (condition III) and approaching the phase of sustained flow the triggered cores are not getting more extended, on the contrary, they are even more localized with $\delta_\mathrm{tgr}\approx2.1$. After unloading (condition IV) the system again has the $\langle v_iv_j\rangle$  behavior seen in the relaxed $Q=10$ state. 

\begin{figure}[ht!]
    \centering
    \includegraphics[width=\columnwidth]{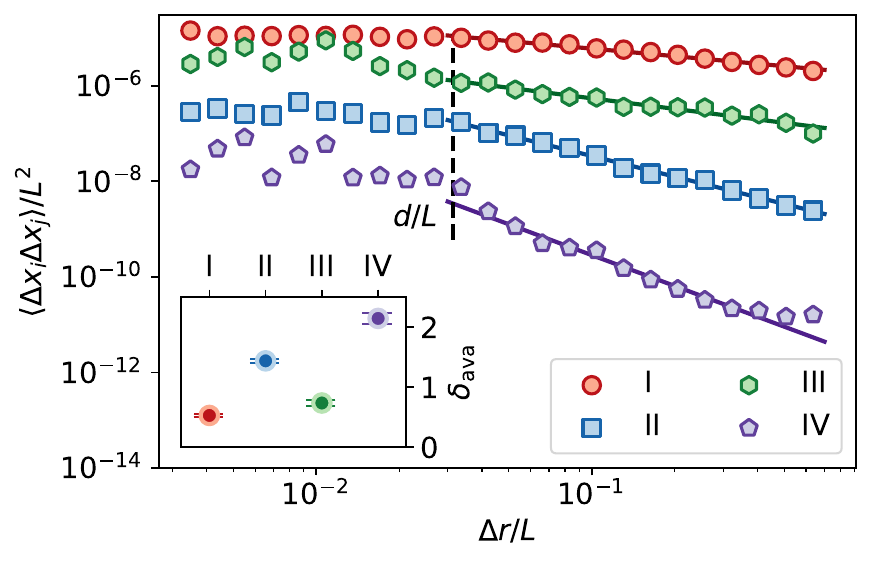}
    \caption{The displacement correlation $\langle \Delta x_i\Delta x_j\rangle$ for displacements of individual dislocations during the first avalanche. The inset show the exponents $\delta_\mathrm{tot}$ defined in Eq. \ref{eq:dxdx_ava_exp}. The dashed line indicates $d/L
$ which denotes the average dislocation spacing normalized with the linear box size $L$.}
    \label{fig:avalanche}
\end{figure}

The displacement correlation $\langle \Delta x_i\Delta x_j \rangle$ is computed based on the displacements during the first avalanche. It also exhibits an asymptotic power-law dependence on $\Delta r$ with an exponent $\delta_\mathrm{ava}$. That is, the asymptotic behavior reads as
\begin{equation}
    \langle \Delta x_i\Delta x_j \rangle(\Delta \bm r)\propto \Delta r ^{-\delta_\mathrm{ava}}.
    \label{eq:dxdx_ava_exp}
\end{equation}
As it clear from Fig. \ref{fig:avalanche} the picture that emerges about the spatial extension of avalanches only partly matches the one seen for the activation cores. Similarly to the case of triggered cores, the avalanches are also quite extended in dislocation systems without quenched disorder (condition I) characterized by an exponent $\delta_\mathrm{ava}\approx0.5$. A significantly higher value of exponent of $\delta_\mathrm{ava}\approx1.4$ at the relaxed $Q=10$ case (condition II) indicates that the introduction of point defects localizes not only the activation cores but the emerging dislocation avalanches as well. This exponent remains high (in fact, even grows) in the prestrained configuration (condition IV) as well. However, for severely strained systems (condition III) $\delta_\mathrm{ava}\approx0.7$ shows that as the system approaches the transition to sustained plastic flow, it starts to behave similarly to the critical $Q=0$ case (condition I) in terms of the spatial extension of the avalanches.

In this paragraph the tendencies in the dynamical correlations are summarized. The exponents $\delta$ of conditions I and II are quite consistent. In the former case the exponents are low ($0<\delta<1$) indicating that the dynamically affected regions of the system are quite extended in space. In the latter case the exponent are significantly higher (around $0.8-0.9$ higher in each case) corresponding to a much more localized dynamic behavior. The strained (loaded) $Q=10$ systems (condition III), however, exhibit a more complex dynamic pattern. Activation cores (where the triggering of the avalanche happens) are still localized (surprisingly even more than in the relaxed systems) but the dynamical correlations in whole avalanches is much longer-ranged (such as in condition I). This is consistent with what one would expect being close to the phase transition to sustained flow since correlations lengths near phase transition tend the diverge (e.g. near the Curie point in the context of magnetization). The prestrained systems (condition IV) have similar dynamical correlations to its relaxed counterpart without any straining history (condition II), in fact, our simulations show that their behavior at the onset of the first avalanche is practically indistinguishable based on distance dependence of $\langle v_iv_j\rangle$. We note that both for the triggering and for the whole avalanches the power-law regime quantified by the exponent $\delta$ was prevalent over the normalized characteristic dislocation spacing $d/L=1/\sqrt{\rho}L=1/\sqrt{N}$. Below $d/L$ typically we obtained roughly constant (distance-independent) velocity and displacement correlations. This indicates that the dislocations that contribute the most to these correlations (i.e. the ones at the very core of the avalanche) typically have comparable velocities irrespective of their actual distance. We also note that our supplementary investigation (not shown here) revealed that changing the (un)loading protocol to gradual (un)loading instead of the instantaneous one presented here do not change the general conclusions of the paper.

\subsection{Participation number}

In the previous section the spatial range of dynamic correlations was examined. In the following the number of dislocations  involved in the (first) avalanches is studied by introducing the participation number (PN). PN characterizes the number of moving dislocations \cite{derlet2016critical} and is defined as 
\begin{equation}
    \mathrm{PN}(v) = \frac{\left(\sum_{i=1}^\mathrm{N}|v_i|^2\right)^2}{\sum_{i=1}^\mathrm{N}|v_i|^4}.
    \label{eq:PN_v}
\end{equation}
Analogously, if the aim is to quantify the dynamic behavior integrated for a time interval instead of an instantaneous dynamic state, the displacement-based version of the participation number can be defined as
\begin{equation}
    \mathrm{PN}(\Delta x) = \frac{\left(\sum_{i=1}^\mathrm{N}|\Delta x_i|^2\right)^2}{\sum_{i=1}^\mathrm{N}|\Delta x_i|^4}.
    \label{eq:PN_dx}
\end{equation}
Here $v_i$ and $\Delta x_i$ are the velocity and the displacement of the $i$th dislocation, respectively. In the special case of $N'$ dislocations moving with the same velocity and $N-N'$ dislocations being still PN simply takes on the value of $N'$. In a general case (potentially all dislocation velocities being different) PN is typically non-integer but stays within the bounds of 1 and $N$ and characterizes the number of dislocations having significant velocities/displacements (compared to the other ones).

\begin{figure}[ht!]
    \centering
    \includegraphics[width=\columnwidth]{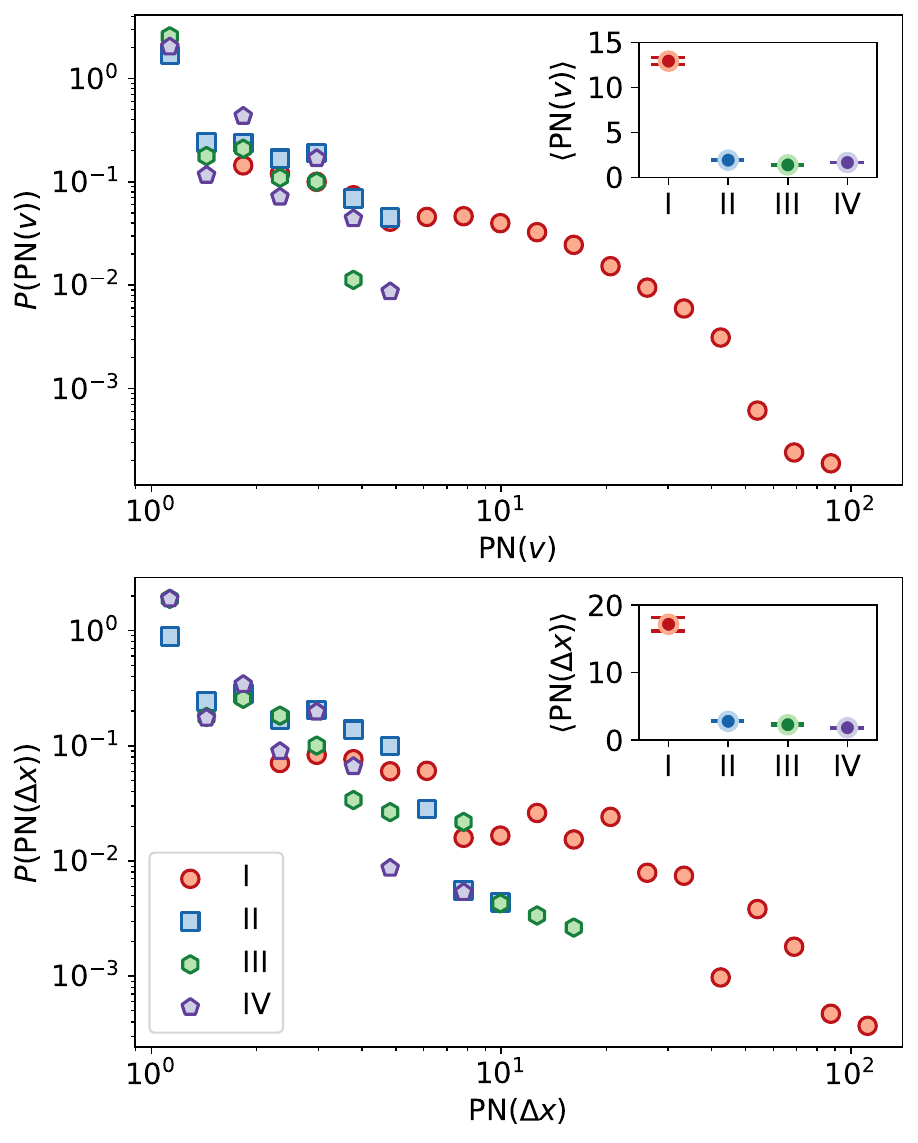}
    \caption{The probability density function $P$ of the participation number PN based on velocities (PN$(v)$) or displacements (PN$(\Delta x)$). For the definition see Eqs. (\ref{eq:PN_v}) and (\ref{eq:PN_dx}). (Top): The PN based on the instantaneous velocities at the onset of the first avalanche. (Bottom): The PN based on the accumulated displacements during the first avalanche. The insets show the mean PN values and their uncertainty estimated with the Jackknife method.}
    \label{fig:PN}
\end{figure}

The probability density functions $P$ of PN$(v)$ and PN$(\Delta x)$ are shown in Fig. \ref{fig:PN}. The results clearly show that more dislocations are involved in both the triggering and the whole avalanche in the $Q=0$ systems (condition I) than in the $Q=10$ ones (conditions II, III and IV) irrespective of the deformation history. It can be observed that different straining history does not yield a striking difference between the average behavior of $Q=10$ systems in terms of the participation number (neither at triggering nor for the whole avalanches). It is somewhat contradictory with the more pronounced differences that were shown in the case of dynamic correlations. In the following section the structure of the avalanches is studied (on the level of individual dislocations) and it is demonstrated that strained configurations (with short-range quenched disorder) can produce avalanches that are spatially quite extended despite having a relatively small number of dislocations involved (which resolves the above mentioned seeming contradiction).  

\subsection{Avalanche structure}

As it was mentioned previously, when comparing the data in terms of participation numbers and spatial dynamic correlations, the results might seem somewhat contradictory in the case of strained $Q=10$ systems (condition III). Namely, while during the avalanches the asymptotic $\langle\Delta x_i \Delta x_j\rangle$ behavior of these systems tends towards the same exponent as for the critical $Q=0$ configurations (condition I), in terms of PN the strained systems are still much closer to the relaxed $Q=10$ case (condition II) which contains much more localized events than condition I. That is, while the avalanches in  strained configurations show longer-range dynamical correlations the number of dislocations involved certainly do no seem to grow proportionally. In order to address this discrepancy, in the following the plastic events are analyzed on the level of individual dislocations.

\begin{figure}[ht!]
    \centering
    \includegraphics[width=\columnwidth]{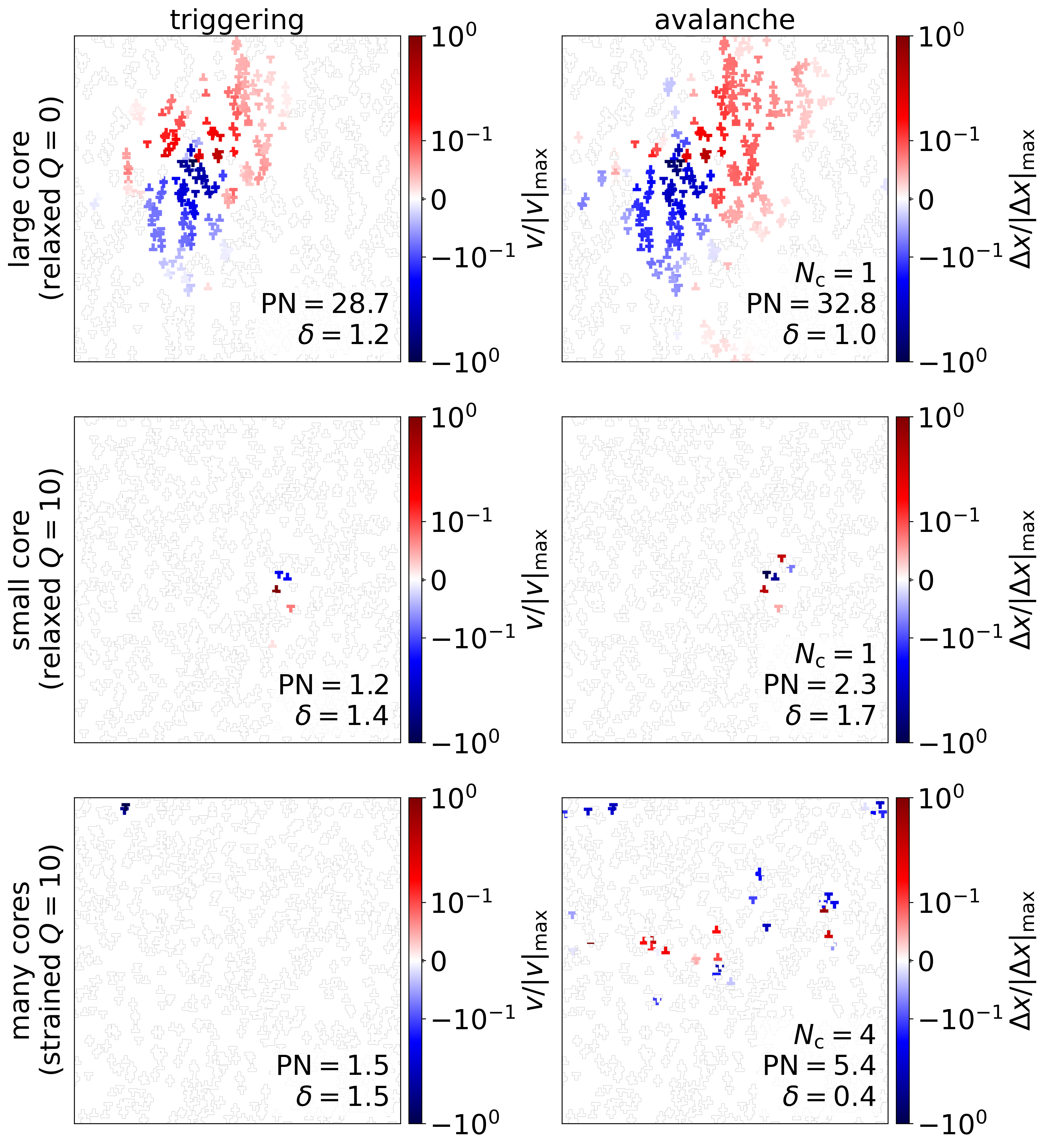}
    \caption{Left: The dislocation velocity $v$ at the triggering at three typical events. Right: The displacement $\Delta x$ accumulated during the same events. The participation number PN (based on either the velocities or the accumulated displacements), the exponent $\delta$ characterizing the decay of either $\langle v_iv_j\rangle$ or $\langle \Delta x_i\Delta x_j\rangle$ at large distances and the number of avalanche cores $N_\mathrm{c}$ are also shown for each event. For the definition of these quantities see the main text. Note: the point defects are not indicated in the figure for visibility purposes.}
    \label{fig:dislocation_level}
\end{figure}

The visual inspection of the avalanches revealed that the events can be categorized into three classes that are presented in Fig. \ref{fig:dislocation_level}. Some events are already quite extended at their onset and naturally the accumulated plastic activity also covers a large portion of the simulation cell (see top row of Fig \ref{fig:dislocation_level}). Other events are triggered very locally and remain small during the avalanche (see middle row). The probably most interesting type of event is the third one where the triggering occurs locally and then small clusters of dislocations (or even single dislocations) are drawn into motion from remote regions of the sample (see bottom row). That is, instead of the growing of a single core of active dislocations a `sea-islands' type picture emerges.

In order to statistically quantify the differences of events in the four conditions a simple algorithm is introduced to detect clusters of active dislocations and classify plastic events into the three casts presented in Fig. \ref{fig:dislocation_level}. To this end, the accumulated dislocation activity during the avalanches is considered. First, all dislocations with a total displacement  below 1\% of the largest individual dislocation displacement are discarded and only the ones with displacements above this threshold are kept. Then all dislocation that are within the distance of $3d$ (where $d=L/\sqrt{N}$ is the average dislocation spacing) are virtually connected. The connected clusters of dislocations are considered the cores of the avalanche. Finally, the number of clusters consisting of more dislocations than the third of the size of the largest core are counted and is denoted by $N_\mathrm{c}$ (for examples see Fig. \ref{fig:dislocation_level}). Additionally, PN$(\Delta x)$ is computed (including all $N$ dislocations). The classes are then determined as follows. If $N_\mathrm{c}>2$, the event is categorized to have `many cores'. If not, that is if it only has 1 or 2 cores, and PN$(\Delta x)>N/100=10.24$ then the event is classified as having a `large core', otherwise, it has a `small core'. We note that while the choice of thresholds for the number of cores and for the participation number is arbitrary, the same conclusions can be drawn within a sensibly wide range of thresholds. 

\begin{figure}[ht!]
    \centering
    \includegraphics[width=\columnwidth]{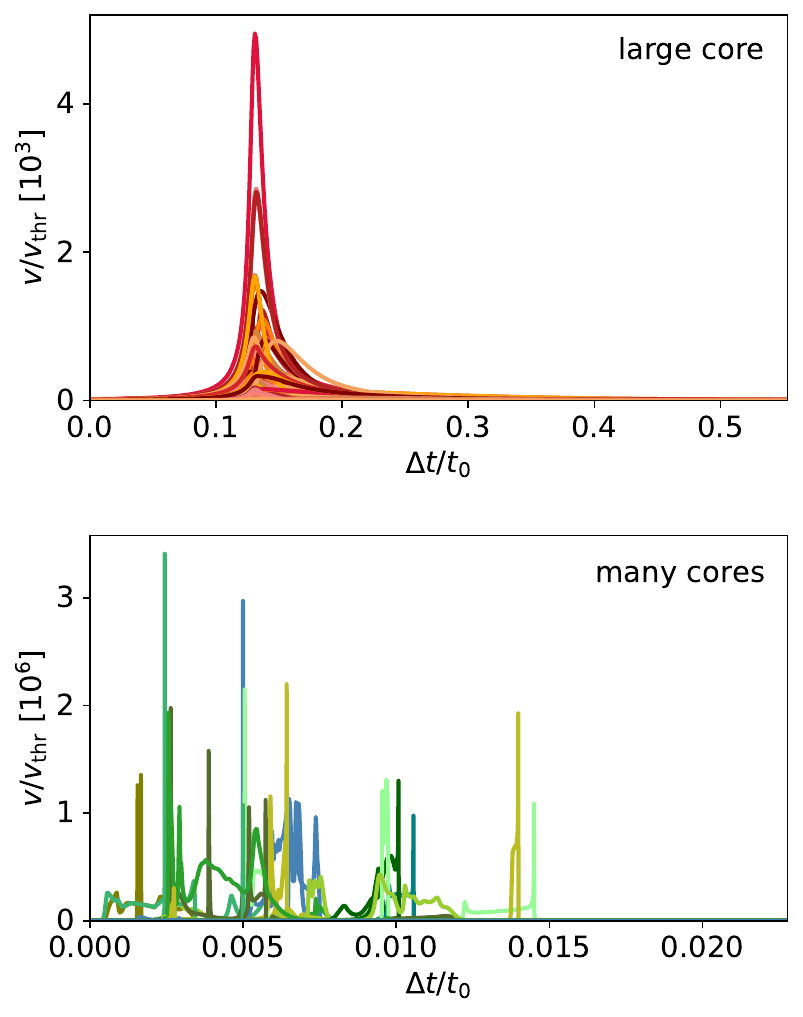}
    \caption{Comparison of the dislocation velocity evolution during avalanches for the two types (single-core and multi-core) of spatially extended plastic events. Each curve denotes the magnitude of the velocity $v$ (normalized with the avalanche detection threshold $v_\mathrm{thr}$) of an individual dislocation. Time $\Delta t$ starts at the onset of the avalanche and is measured in simulation unit $t_0$. (Top): A typical single-core event (corresponding to the `large core' example in Fig. \ref{fig:dislocation_level}) which has a relatively smooth and uniform velocity evolution. (Bottom): A typical multi-core avalanche (corresponding to the `many cores' example in Fig. \ref{fig:dislocation_level}) characterized by the chain-triggering of remote dislocations (or clusters of dislocations) resulting in an intermittency of the evolution of dislocation velocity.}
    \label{fig:chain_triggering}
\end{figure}

\begin{figure}[ht!]
    \centering
    \includegraphics[width=0.75\columnwidth]{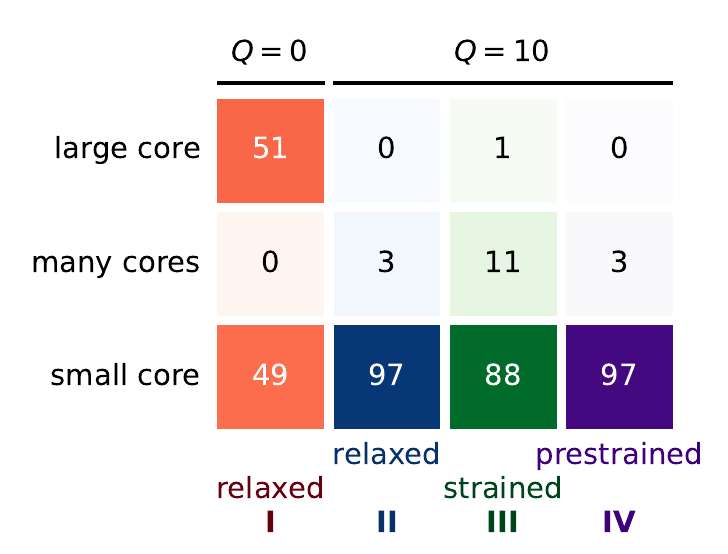}
    \caption{The classification of the first avalanches of the 100 configurations of each type. Relaxed $Q=0$ systems exhibit a diversity between of small and large typically single-core avalanches. Relaxed and prestrained $Q=10$ configurations almost exclusively produce small single-core avalanches while a significant minority of multi-core but typically relatively low-PN events emerges in strained systems. For the details of the classification method and the definition of the three classes see the main text.}
    \label{fig:event_type}
\end{figure}

The execution of this classification on the four conditions reveals the following picture that is visualized in Fig. \ref{fig:event_type}. As it was shown, the behavior of relaxed $Q=0$ and $Q=10$ systems (conditions I and II) is fundamentally different. In $Q=10$ configurations practically all plastic events are quite small while in the $Q=0$ samples their is a wide selection of events ranging from the motion of single dipoles ($PN\approx2$) up to events  well above our threshold of $N/100=10.24$ (see also Fig. \ref{fig:PN}). It is the same, however, in both cases that the core of the avalanche is typically one spatially compact active region. The picture is more complex as the $Q=10$ configuration is strained and approaches the phase transition to the state of sustained flow (condition III). While the growth in terms of PN is incremental, the activation of many small and spatially disjoint parts of the sample during the avalanche is significantly (c. four times) more frequent than in the relaxed case. Individual dislocation velocity data suggests that remote dislocation clusters are drawn into the avalanche by subsequent chain-triggering (see Fig. \ref{fig:chain_triggering} for an example). This explains the conundrum how the long-range dynamical correlations appear in the strained systems even though the size of the avalanches (in terms of the number of participating dislocations) does not increase proportionally at all.  It can be also observed that upon unloading $Q=10$ systems (condition IV) return to the same behavior characterizing their relaxed counterparts (condition II) despite their pre-deformation history.

\begin{figure}[ht!]
    \centering
    \includegraphics[width=\columnwidth]{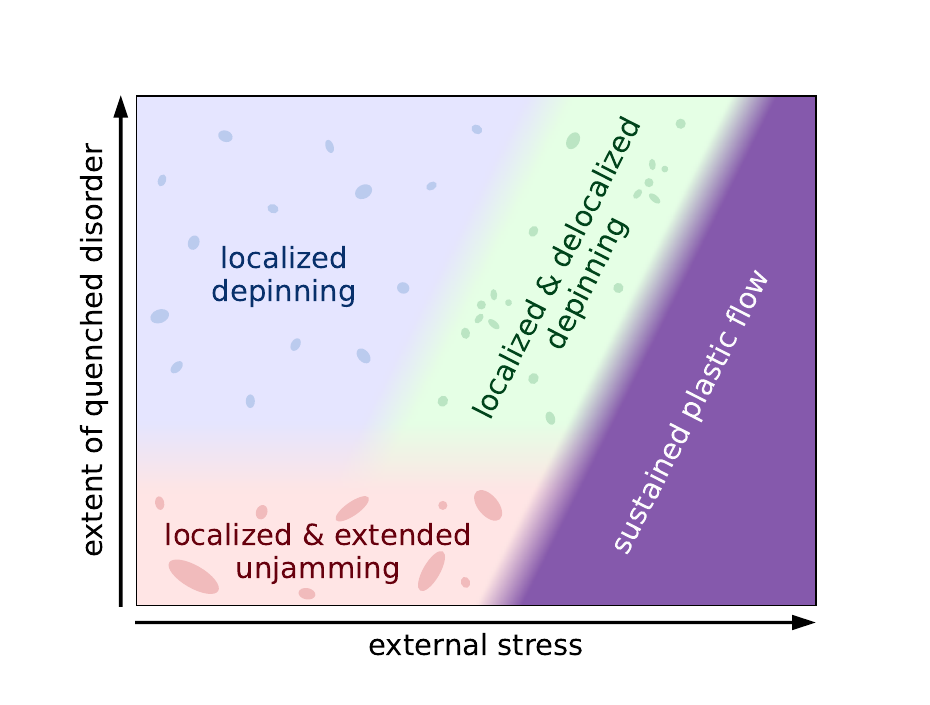}
    \caption{The regimes of dynamic behavior during avalanche activity. At low external stress in systems dominated by long-range dislocation-dislocation interactions (that is, with no or small extent of quenched disorder) avalanches occur due to the unjamming of dislocations in form of compact events that come from a very wide size-range. At low external stress in systems dominated by short-range interactions (that is, in the presence of large extent of quenched disorder) upon depinning small localized avalanches occur. At larger stresses however, a significant minority of delocalized avalanches appear in these systems which avalanches consist of small clusters of moving dislocation scattered across a large area. At large enough stress (which is dependent on the extent of short-ranged quenched disorder) the system reaches the state of sustained plastic flow.}
    \label{fig:phase-space-new}
\end{figure}

It was demonstrated in previous works that at small strains pure (point defect free) dislocation systems (condition I) and the ones also containing a great extent of short-range quenched disorder (in the form of, e.g., point defects or precipitates, condition II) belong to two distinct classes \cite{ispanovity2014avalanches, ovaska2015quenched}. This has been observed here as well: in both conditions I and II the avalanches have a single compact core, however, while in condition I localized and extended events coexists, in condition II practically all avalanches are localized. The current results imply that the avalanche behavior of immensely strained systems with quenched disorder (condition III) belong to a third, different regime. In these systems there is diversity of events affecting very local and very extended regions of the sample, but even in the latter case the avalanches consist of a small number of dislocations which are, however, from clusters scattered along the specimen. The proposed phase diagram supplemented with the new regime is schematically shown in Fig. \ref{fig:phase-space-new} with caption briefly summarizing the nature of avalanches in each regime. It should be stressed that the figure is strictly schematic and the actual shapes of different regimes can not be obtained from the current data.

\subsection{Yield stresses and local yield stresses}

After studying the correlations and the avalanche behavior now we turn to one of the main questions if this work: how are the local yield stresses affected by the avalanches and the deformation of the model sample. Firstly, the long-term, global impact of the deformation is studied. That is, the effect of severely straining and then unloading the system on the statistics on the yield stresses. Then, the short-term, local impact is investigated. Namely, how the local yield stress landscape evolves due to individual avalanches in the case of the four different conditions. 

\begin{figure}
    \centering
    \includegraphics[width=\columnwidth]{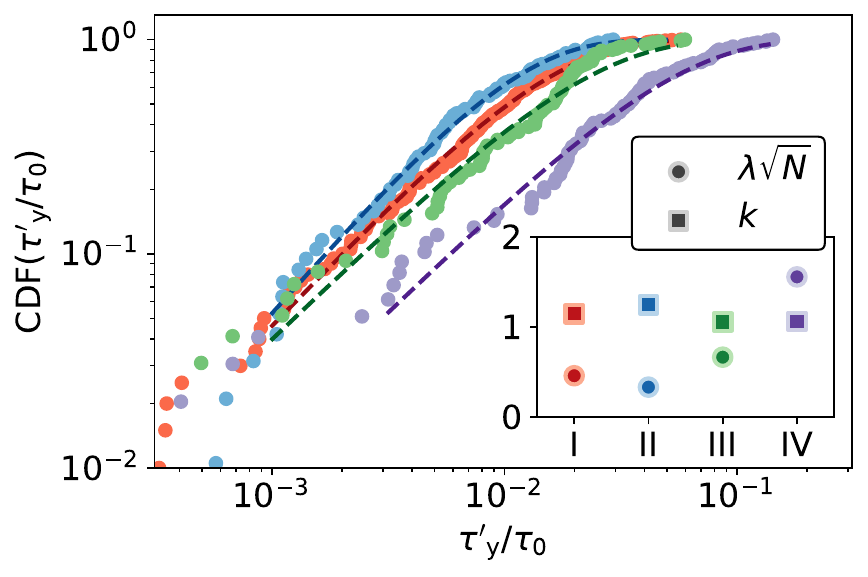}
    \caption{The cumulative distribution functions (CDFs) of the corrected yield stress $\tau'_\mathrm{y}=\tau_\mathrm{y}-\tau_\mathrm{init}$ where $\tau_\mathrm{y}$ and $\tau_\mathrm{init}$ are the yield stress and the initial external stress, respectively. The latter is non-zero only in the strained (green) case. The inset shows the scale ($k$) and shape ($\lambda$) parameters of the fitted Weibull distributions. $\tau_0$ is a constant defined the main text and $N=1024$ is the number of dislocations. }
    \label{fig:lys_cdf}
\end{figure}

The local yield stress associated with a given sub-region of the sample (referred to as \emph{box} in the rest of the paper) is assumed to be determined by the strength of the weakest of the activable substructures (links) within the box. (If the local yield stress is assigned to the whole simulation cell, we will simply refer to it as the yield stress or yield threshold). Thus, one may expect the local yield stress $\tau_\mathrm{y}$ to obey an extremal probability distribution, in particular, a first order statistics (i.e., minimum) distribution. Namely, if in the small strength limit the link strength $\tau_\mathrm{link}$ is power-law distributed as $F_\text{link}(\tau_\text{link}) \propto \tau_\text{link}^k$, $F_\text{link}$ being the cumulative distribution function (CDF) of $\tau_\mathrm{link}$, then the emergent extremal distribution is of Weibull type \cite{weibull1939statistical, weibull1951wide, derlet2015probabilistic, derlet2015universal, ispanovity2017role} with a CDF
\begin{equation}
    F(\tau_\mathrm{y})=1-\mathrm{exp}\left[-\left(\frac{\tau_\mathrm{y}}{\lambda}\right)^k\right].
    \label{eq:weibull}
\end{equation}
Here $\lambda$ and $k$ are the so-called \emph{scale} and \emph{shape parameters}, respectively. The former is proportional to the mean value of the variable and the latter describes the asymptotic behavior at the $\tau_\mathrm{y}\rightarrow{}0$ limit. Here we consider $\tau'_\mathrm{y}=\tau_\mathrm{y}-\tau_\mathrm{init}$ where $\tau_\mathrm{init}$ is the initial external stress which is zero for all conditions except condition III. The statistics of  yield thresholds $\tau'_\mathrm{y}$ (corresponding to the whole simulation cell) were obtained during loading the sample with a quasi-statically increasing spatially homogeneous external stress, $\tau_\mathrm{y}$ being the critical stress at which the first avalanche occurs. The event is detected based on thresholding: the avalanche is considered to have started when both the mean dislocation velocity and its derivative w.r.t time exceed a predefined value. As it is shown in Fig. \ref{fig:lys_cdf} the statistics of the yield thresholds $\tau'_\mathrm{y}$ indeed obeys the Weibull distribution with a shape parameter $k\approx1.1\pm0.1$ robust to the change of the extent $Q$ of quenched disorder and to the deformation history of the sample. The scale parameter $\lambda$, however, depends on $Q$ and the deformation history. The most significant effect is that prestraining (condition IV) immensely increases the yield stress values due to the elimination of weak substructures that are easy to trigger during the loading stage.

\begin{figure}[ht!]
    \centering
    \includegraphics[width=\columnwidth]{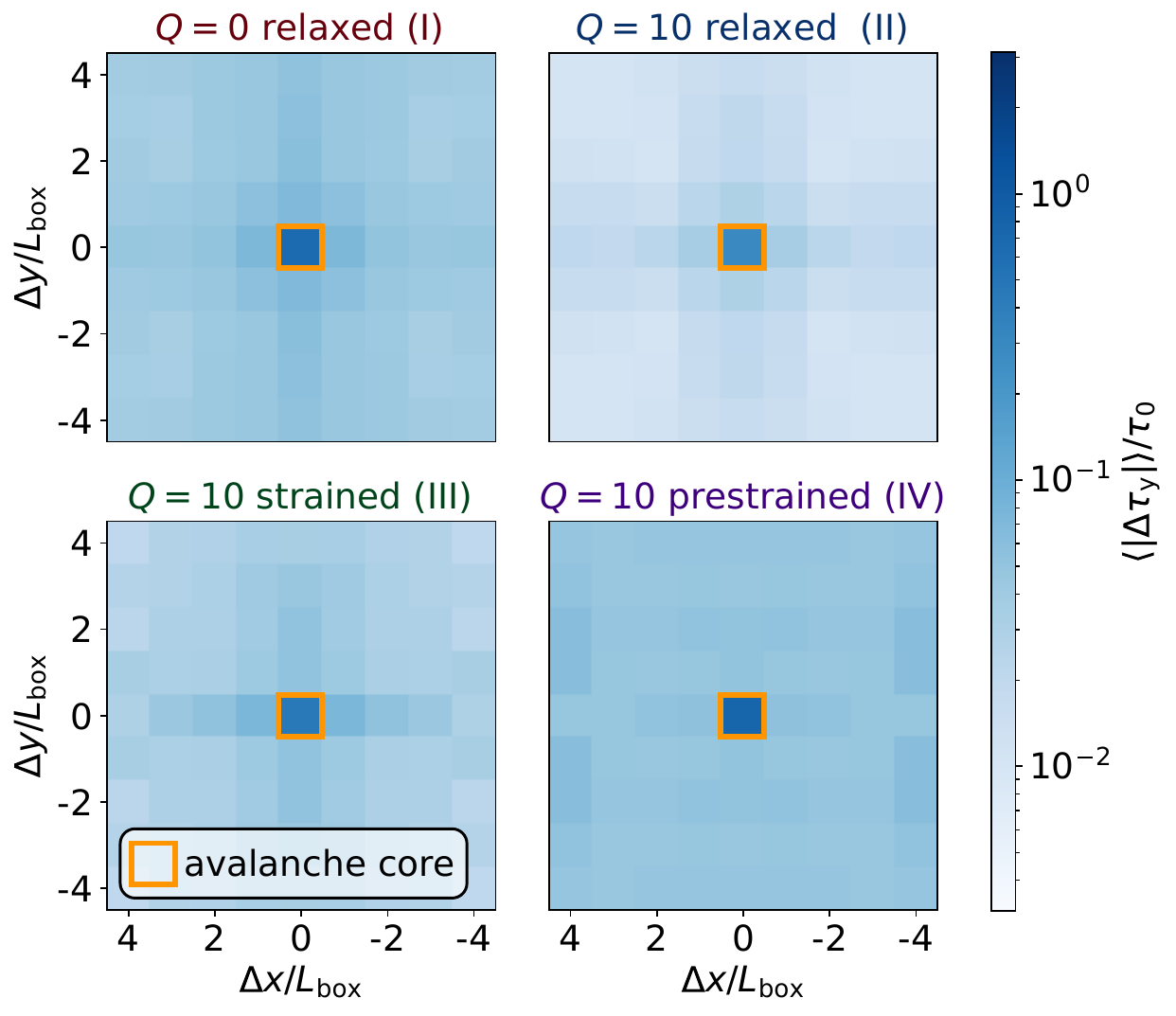}
    \caption{The average of the absolute value of the change of the local yield stresses $\langle |\Delta \tau_\mathrm{y}|\rangle$ in the boxes. $\Delta x$ and $\Delta y$ are the relative coordinates measured from the middle of the box where $|\Delta \tau_\mathrm{y}|$ is the largest. $L_\mathrm{box}$ is the edge width of the boxes and $\tau_0$ is a constant defined in the main text.}
    \label{fig:dlys_abs_map}
\end{figure}

\begin{figure}[ht!]
    \centering
    \includegraphics[width=\columnwidth]{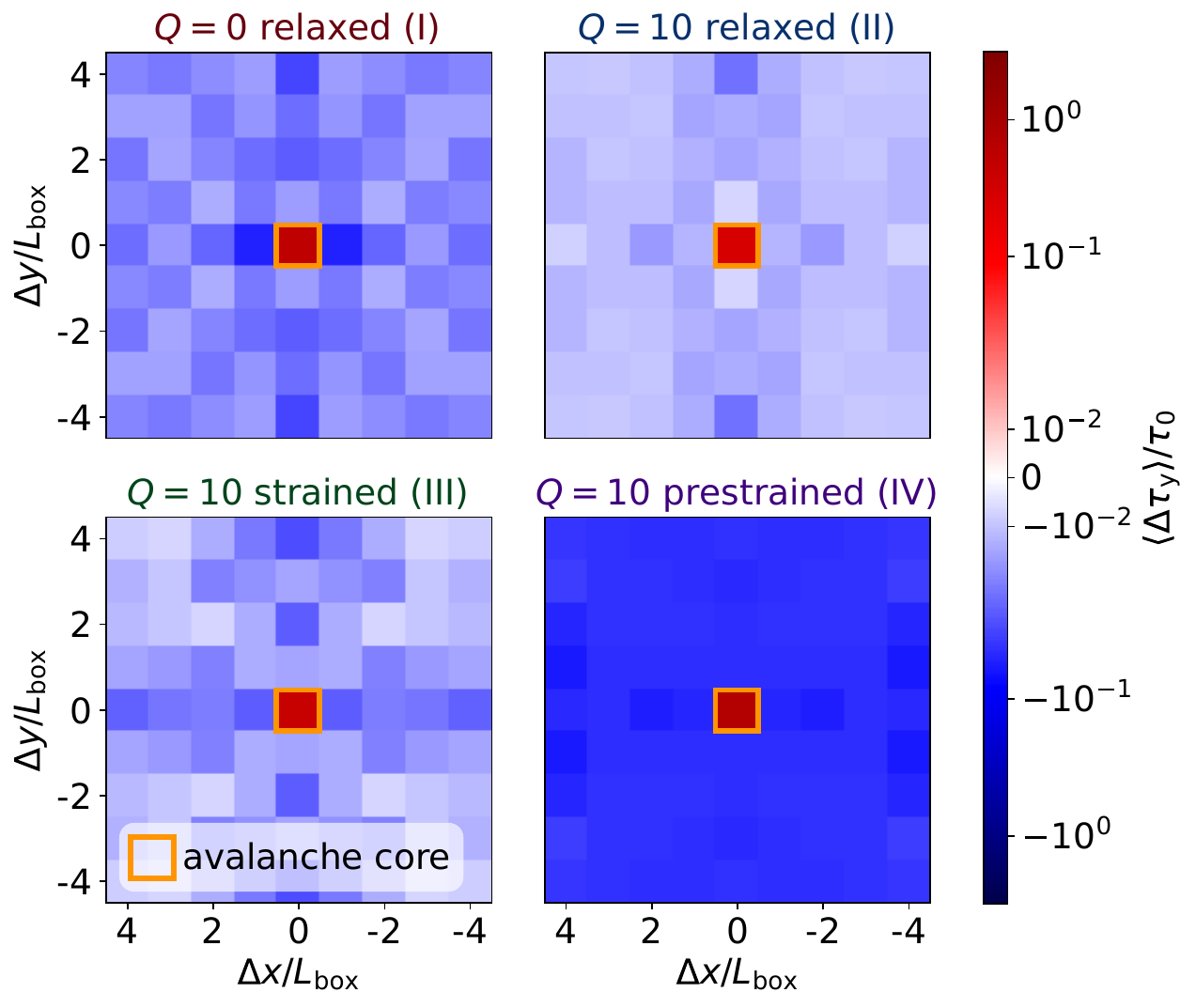}
    \caption{The average of the change of the local yield stresses $\langle \Delta \tau_\mathrm{y}\rangle$ in the boxes. $\Delta x$ and $\Delta y$ are the relative coordinates measured from the middle of the box where $|\Delta \tau_\mathrm{y}|$ is the largest. $L_\mathrm{box}$ is the edge width of the boxes and $\tau_0$ is a constant defined in the main text. The middle box (where the avalanche core is) exhibits a significant hardening while the surrounding boxes show softening on average.}
    \label{fig:dlys_map}
\end{figure}

Above it was shown that while some features (e.g. the shape parameter) of the yield threshold distribution is robust, the straining history can affect the statistics of local yield stresses. Now let us focus on the short-term changes of local yield stresses (corresponding to smaller subsystems) after individual dislocation avalanches. To this end the following procedure is applied. Each configuration is divided into $8\times8$ square shaped disjoint boxes. Then the local yield stress of each box is probed by quasi-statically increasing spatially homogeneous external stress while keeping the dislocations outside the box fixed. The computation of local yield stresses is done with the thresholding-based detection of avalanches described above. The local yield stresses are first computed for all boxes based on the first avalanches. Then,  quasi-static global loading is applied to the relaxed configurations. That is, all dislocations are affected by the homogeneous external stress and all are mobile. The configurations are loaded until the first avalanche occurs (that is, until the yield threshold of the whole simulation cell). This, naturally results in the spatial redistribution of dislocations which leads to the change of the local yield stress of each fictitious box. The local yield stresses are measured after this global loading as well.  The change of $\tau_\mathrm{y}$ is denoted with $\Delta\tau_\mathrm{y}$. For each configuration a map of $\Delta\tau_\mathrm{y}$ is created. The box with the largest change (in terms of absolute value) is considered to be the core of the avalanche was. Each map is rearranged by moving the core box into the center while keeping the relative positions (quantified by coordinates $\Delta x$ and $\Delta y$) of boxes (taking into account the PBC). These re-centered maps are than averaged (and symmetrized in order to decrease noise) over the ensembles of configurations of each of the four conditions. The so obtained maps of $|\Delta \tau_\mathrm{y}|$ and $\Delta\tau_\mathrm{y}$ are shown in Figs. \ref{fig:dlys_abs_map} and \ref{fig:dlys_map}, respectively. 

\begin{figure}[ht!]
    \centering
    \includegraphics[width=\columnwidth]{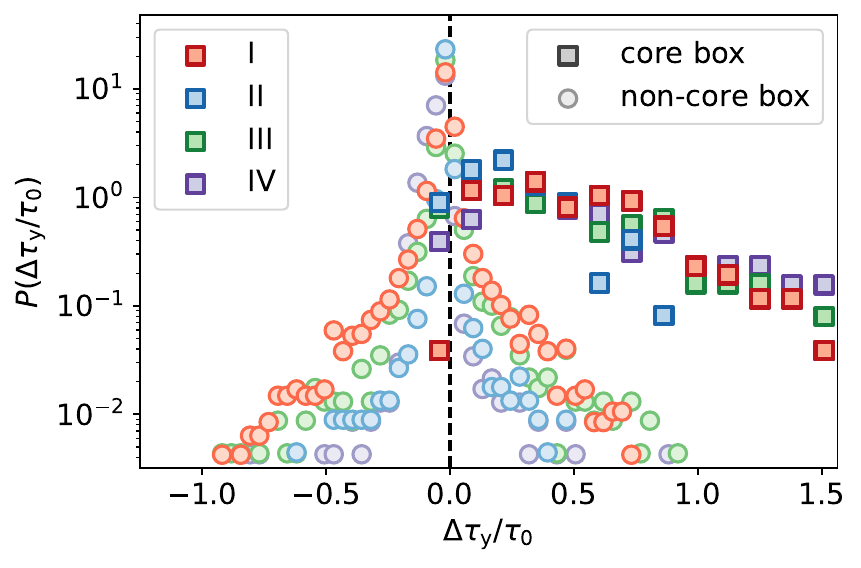}
    \caption{The probability density function $P$ of the change of local yield stresses $\Delta\tau_\mathrm{y}$ in core boxes (where the core of the avalanche is) and non-core boxes. The core boxes almost always exhibit strong hardening and the non-core ones typically show weak softening. $\tau_0$ is a constant defined in the main text and the dashed line underscores $\Delta \tau_\mathrm{y}=0$.}
    \label{fig:dlys_pdf}
\end{figure}

The maps indicate that at the core box (that is, where the dislocation avalanche emerged) $\langle|\Delta\tau_\mathrm{y}|\rangle$ is significantly larger than in any other box and $\langle\Delta\tau_\mathrm{y}\rangle>0$, that is, an immense hardening is observed. This is not surprising since after the activation of the first avalanche the next activable link is probably significantly harder to trigger. This can be explained by the assumption (consistent with the numerically obtained Weibull distributions of $\tau_\mathrm{y}$) that the low-strength asymptotic distribution of link strengths obeys a power-law. This is the part of the distribution that matters during weakest link activation. Since at the low-strength tail of the distribution the probability density is low, independently of the redistribution of dislocations, one's intuition would be that the second avalanche in this core box is significantly harder to trigger than the first one. The other boxes are also affected by the avalanche but to a smaller extent and typically softening can be observed instead of hardening. The probability density functions of $\Delta\tau_\mathrm{y}$ are shown in Fig.~\ref{fig:dlys_pdf}. The core boxes almost always (with more than 93\% probability in each condition) harden with $\Delta\tau_\mathrm{y}$ approximately obeying an exponential distribution. The non-core boxes typically soften (with more than 76\% probability in each condition) due to the  increased external stress and the change of internal stresses resulted by the dislocation rearrangements.

\begin{figure}
    \centering
    \includegraphics[width=\columnwidth]{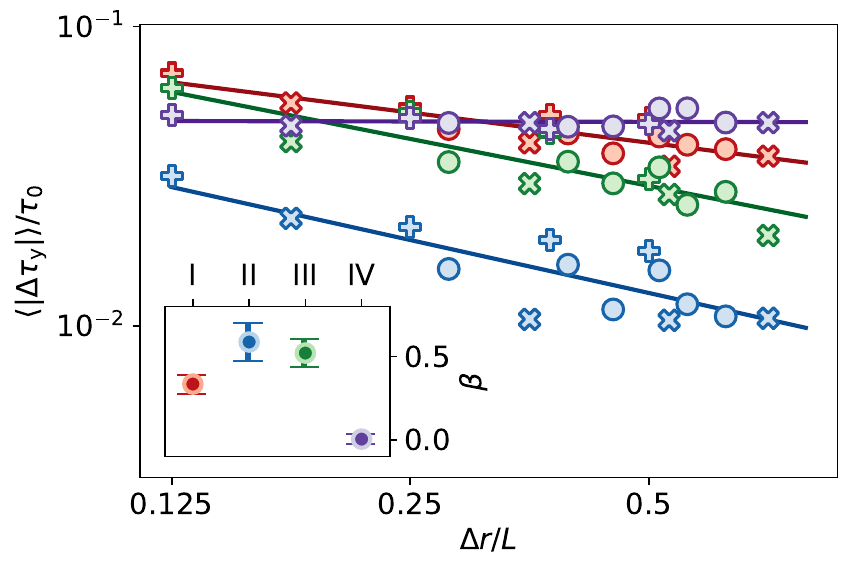}
    \caption{The average of the absolute value of the change of the local yield stresses $\langle |\Delta \tau_\mathrm{y}|\rangle$ in the boxes as a function of the distance $\Delta r$ of their middle from the middle of the box that contains the core of the avalanche (that is, where $|\Delta \tau_\mathrm{y}|$ is maximal). `$+$': box horizontally or vertically aligned with the core box, `$\times$': box diagonally positioned in comparison to the core box, `$\circ$': every other box. The inset shows the exponent $\beta$ characterizing the power-law decay with the distance from the core box. $L$ and $\tau_0$ are the linear size of the simulation cell and a constant defined in the main text, respectively.}
    \label{fig:dlys_radial}
\end{figure}

Figure \ref{fig:dlys_radial} shows the dependence of $\langle|\Delta \tau_\mathrm{y}|\rangle$ on the distance $\Delta r = \sqrt{\Delta x + \Delta y}$ measured from the avalanche core. In every condition the impact of the avalanche on the local yield stresses decay according to
\begin{equation}
    \langle|\Delta \tau_\mathrm{y}|\rangle \propto (\Delta r)^{-\beta}.
\end{equation}
As in relaxed $Q=10$ systems (condition II) the dynamical correlations are weaker and shorter-ranged than in their $Q=0$ counterparts (condition I), not surprisingly, the same can be observed in terms of the magnitude of $\langle|\Delta \tau_\mathrm{y}|\rangle$ and the exponent $\beta$. More surprisingly, $\beta$ remain the same during severe straining (condition III), although, the avalanche impact is stronger (comparable with condition I) indicated by the increase in the magnitude of $\langle|\Delta \tau_\mathrm{y}|\rangle$. In the prestraining case (condition IV) $\beta\approx0$, that is, the softening is comparable in all boxes. This is the results of the first avalanche being triggered after an increment of external stress much larger than in the case of any other conditions which leads to a larger extent of rearrangement (in the form of drifting dislocation motion) prior to the avalanche. This is probably related to the elimination of substructures that are easy to trigger during the prestraining (which was also suggested in Ref. \cite{salmenjoki2018machine}). 

It is generally true (regardless of the condition) that the boxes that are vertically or horizontally aligned with the core box are more impacted by the avalanche while the diagonally positioned boxes are less affected. This angular dependence can be related to the stress field induced by the rearrangement of the dislocations by the inspection of the Eshelby stress field of a displaced dislocation \cite{eshelby1957determination}. The shear stress field of a dislocation displaced with an infinitesimally small $\Delta x$ changes with $\Delta \tau=\partial_x \tau_d\Delta x=\frac{\cos(4\varphi)}{r^2}\Delta x$ where $\varphi$ is the angular polar coordinate and $\tau_\mathrm{d}$ is the shear stress field of a dislocation (see Eq. (\ref{eq:stress_field})). If it is assumed that (i) during an avalanche the displacement of dislocations is typically small, (ii) most of the dislocations move in the direction consistent with the external stress and (iii) the dislocation motion is the most intense in a relatively local and compact inner core of the avalanche, the change of the internal stress field has roughly a $\cos(4\varphi)$ type angular dependence (consistently with the results in Ref. \cite{ispanovity2014avalanches}) which strengthens the effect of the external stress in the vertical and horizontal directions and weakens it diagonally. This leads to the observed enhanced softening vertically and horizontally and the mitigated softening diagonally.

We note that the observations above are not limited to the box size corresponding to the resolution of $8\times8$. Our prior investigation revealed that for finer resolutions than $8\times8$ empty boxes may appear which formally have infinite local yield stress in our model. Therefore, the resolution of $8\times8$ was chosen since it is suitable for our analysis and provides a more detailed picture than coarser resolutions. However, it should be noted that our analyses (not shown here) conducted at resolutions $4\times4$ and $6\times6$ corroborates that the general conclusions drawn above (e.g., the $\cos(4\varphi)$-type anisotropy, the approximately power-law decay of $|\Delta\tau_\mathrm{y}|$ and the differences between the four conditions) do not depend on the actual box size.

\section{Conclusion and outlook}
In this work we studied the static and dynamic length scales, avalanche dynamics and local yield stresses in 2D DDD framework in configurations with and without short-range quenched disorder and with different deformation history. As the pure systems driven exclusively by long-range elastic interactions  exhibit criticality already at zero external stress, our focus was on whether and how the behavior of systems with quenched disorder (dominated by short-range interactions) tend to the criticality characterizing the pure systems. Therefore, while using the relaxed systems (with or without short-ranged quenched disorder) as references we examined the behavior of systems with quenched disorder after severely straining (loading) them and also after subsequent unloading.

In terms of static dislocation-dislocation correlations it was found that the straining, despite making the correlation weaker, also makes it longer-range which effect persisted after unloading the sample as well. Similarly, the severely strained systems (despite  avalanches being triggered very locally, in contrary to pure systems) exhibited longer-range dynamic correlations than their relaxed counterparts (unlike the unloaded/prestrained configurations). However, further analysis of the data revealed that this avalanche behavior, despite exhibiting similar asymptotic dynamic correlations compared to pure systems, fundamentally differs from the nature of plastic events in the systems dominated by long-range interactions. Our results imply that apart from the regime of sustained plastic flow at very large stresses at least three distinct regimes of avalanche behavior exist in this framework. Namely, (1) when long-range dislocation-dislocation interaction dominate the dynamics, localized and extended plastic events coexist and they occur as a result of dislocation unjamming. In systems dominated by short-range interactions (2) at low external stress plastic events are localized and happen due to depinning of dislocations, however, (3) at larger stresses, in addition to localized events, another type of avalanche appears. These events are typically also triggered locally but the triggering is followed by a delocalized sequence of depinning of small, spatially isolated and often quite remote clusters of dislocations. That is, while in this regime the extended dynamic correlations indeed appear similarly to the pure systems these correlations are not the result of the correlated motion of a large and compact dislocation cluster but the chain-triggering of a lower number of dislocations scattered across a quite extended region of the system.

Besides the correlations, the yield stress (corresponding to whole configurations) and local yield stresses (corresponding to subsystems) were studied. The results showed that the yield stresses (consistently with the weakest-link principle) obey Weibull distribution with a shape parameter robust to the introduction of quenched disorder and to the deformation history. It was observable, however, that the loading and subsequent unloading of a configuration results in a remarkable hardening (that is, a shift of the scale parameter of the yield stress distribution). The analysis of local yield stresses indicate that the local hardness of the material changes significantly when an avalanche occurs due to the rearrangement of dislocations. At the core of the avalanche an immense hardening is observable. In the more remote vicinity, however, the material is (on average) softening and the impact is significantly weaker. The softening effect diminishes as the distance from the core of the avalanche is increased. More precisely, the mean change of the local yield stress decays according to power-law with an exponent dependent on the extent of quenched disorder and the straining history  and it exhibits an angular dependence consistent with the Eshelby stress field.

We note that our results were performed in a relatively simple 2D DDD model which as such can reproduce statistical features of crystalline plasticity in single-slip setup (such as in appropriately oriented HCP materials) but it cannot account for more complex behavior prevalent in multi-slip scenarios, e.g., junctions, cross-slip, etc.  It still remains to be challenging, however, to study the high-strain regime for sufficiently large samples in the 3D DDD framework due to the immense computational cost. This computational barrier and the easier interpretability of the 2D model led us to employing 2D DDD for this particular study. Nevertheless, it would be an interesting question in the future to see how severe straining affects the static and dynamic correlations and local yield stresses in more complex 3D DDD models and whether a similar picture emerges in terms of the prevalent regimes of avalanche behavior.

\begin{acknowledgments}
Support by the National Research, Development and Innovation Fund of Hungary (contract numbers: NKFIH-FK-138975) is acknowledged.
Supported by the ÚNKP-21-1 New National Excellence Program of the Ministry for Innovation and Technology from the source of the National Research, Development and Innovation Fund. 
\end{acknowledgments}

\bibliography{weakest}

\begin{thebibliography}{40}%
\makeatletter
\providecommand \@ifxundefined [1]{%
 \@ifx{#1\undefined}
}%
\providecommand \@ifnum [1]{%
 \ifnum #1\expandafter \@firstoftwo
 \else \expandafter \@secondoftwo
 \fi
}%
\providecommand \@ifx [1]{%
 \ifx #1\expandafter \@firstoftwo
 \else \expandafter \@secondoftwo
 \fi
}%
\providecommand \natexlab [1]{#1}%
\providecommand \enquote  [1]{``#1''}%
\providecommand \bibnamefont  [1]{#1}%
\providecommand \bibfnamefont [1]{#1}%
\providecommand \citenamefont [1]{#1}%
\providecommand \href@noop [0]{\@secondoftwo}%
\providecommand \href [0]{\begingroup \@sanitize@url \@href}%
\providecommand \@href[1]{\@@startlink{#1}\@@href}%
\providecommand \@@href[1]{\endgroup#1\@@endlink}%
\providecommand \@sanitize@url [0]{\catcode `\\12\catcode `\$12\catcode
  `\&12\catcode `\#12\catcode `\^12\catcode `\_12\catcode `\%12\relax}%
\providecommand \@@startlink[1]{}%
\providecommand \@@endlink[0]{}%
\providecommand \url  [0]{\begingroup\@sanitize@url \@url }%
\providecommand \@url [1]{\endgroup\@href {#1}{\urlprefix }}%
\providecommand \urlprefix  [0]{URL }%
\providecommand \Eprint [0]{\href }%
\providecommand \doibase [0]{http://dx.doi.org/}%
\providecommand \selectlanguage [0]{\@gobble}%
\providecommand \bibinfo  [0]{\@secondoftwo}%
\providecommand \bibfield  [0]{\@secondoftwo}%
\providecommand \translation [1]{[#1]}%
\providecommand \BibitemOpen [0]{}%
\providecommand \bibitemStop [0]{}%
\providecommand \bibitemNoStop [0]{.\EOS\space}%
\providecommand \EOS [0]{\spacefactor3000\relax}%
\providecommand \BibitemShut  [1]{\csname bibitem#1\endcsname}%
\let\auto@bib@innerbib\@empty
\bibitem [{\citenamefont {Fleck}\ \emph {et~al.}(1994)\citenamefont {Fleck},
  \citenamefont {Muller}, \citenamefont {Ashby},\ and\ \citenamefont
  {Hutchinson}}]{fleck1994strain}%
  \BibitemOpen
  \bibfield  {author} {\bibinfo {author} {\bibfnamefont {N.}~\bibnamefont
  {Fleck}}, \bibinfo {author} {\bibfnamefont {G.}~\bibnamefont {Muller}},
  \bibinfo {author} {\bibfnamefont {M.}~\bibnamefont {Ashby}}, \ and\ \bibinfo
  {author} {\bibfnamefont {J.}~\bibnamefont {Hutchinson}},\ }\href@noop {}
  {\bibfield  {journal} {\bibinfo  {journal} {Acta Metallurgica et materialia}\
  }\textbf {\bibinfo {volume} {42}},\ \bibinfo {pages} {475} (\bibinfo {year}
  {1994})}\BibitemShut {NoStop}%
\bibitem [{\citenamefont {Uchic}\ \emph {et~al.}(2004)\citenamefont {Uchic},
  \citenamefont {Dimiduk}, \citenamefont {Florando},\ and\ \citenamefont
  {Nix}}]{uchic2004sample}%
  \BibitemOpen
  \bibfield  {author} {\bibinfo {author} {\bibfnamefont {M.~D.}\ \bibnamefont
  {Uchic}}, \bibinfo {author} {\bibfnamefont {D.~M.}\ \bibnamefont {Dimiduk}},
  \bibinfo {author} {\bibfnamefont {J.~N.}\ \bibnamefont {Florando}}, \ and\
  \bibinfo {author} {\bibfnamefont {W.~D.}\ \bibnamefont {Nix}},\ }\href@noop
  {} {\bibfield  {journal} {\bibinfo  {journal} {Science}\ }\textbf {\bibinfo
  {volume} {305}},\ \bibinfo {pages} {986} (\bibinfo {year}
  {2004})}\BibitemShut {NoStop}%
\bibitem [{\citenamefont {Uchic}\ \emph {et~al.}(2009)\citenamefont {Uchic},
  \citenamefont {Shade},\ and\ \citenamefont {Dimiduk}}]{uchic2009plasticity}%
  \BibitemOpen
  \bibfield  {author} {\bibinfo {author} {\bibfnamefont {M.~D.}\ \bibnamefont
  {Uchic}}, \bibinfo {author} {\bibfnamefont {P.~A.}\ \bibnamefont {Shade}}, \
  and\ \bibinfo {author} {\bibfnamefont {D.~M.}\ \bibnamefont {Dimiduk}},\
  }\href@noop {} {\bibfield  {journal} {\bibinfo  {journal} {Annual Review of
  Materials Research}\ }\textbf {\bibinfo {volume} {39}},\ \bibinfo {pages}
  {361} (\bibinfo {year} {2009})}\BibitemShut {NoStop}%
\bibitem [{\citenamefont {Dimiduk}\ \emph {et~al.}(2006)\citenamefont
  {Dimiduk}, \citenamefont {Woodward}, \citenamefont {LeSar},\ and\
  \citenamefont {Uchic}}]{dimiduk2006scale}%
  \BibitemOpen
  \bibfield  {author} {\bibinfo {author} {\bibfnamefont {D.~M.}\ \bibnamefont
  {Dimiduk}}, \bibinfo {author} {\bibfnamefont {C.}~\bibnamefont {Woodward}},
  \bibinfo {author} {\bibfnamefont {R.}~\bibnamefont {LeSar}}, \ and\ \bibinfo
  {author} {\bibfnamefont {M.~D.}\ \bibnamefont {Uchic}},\ }\href@noop {}
  {\bibfield  {journal} {\bibinfo  {journal} {Science}\ }\textbf {\bibinfo
  {volume} {312}},\ \bibinfo {pages} {1188} (\bibinfo {year}
  {2006})}\BibitemShut {NoStop}%
\bibitem [{\citenamefont {Csikor}\ \emph {et~al.}(2007)\citenamefont {Csikor},
  \citenamefont {Motz}, \citenamefont {Weygand}, \citenamefont {Zaiser},\ and\
  \citenamefont {Zapperi}}]{csikor2007dislocation}%
  \BibitemOpen
  \bibfield  {author} {\bibinfo {author} {\bibfnamefont {F.~F.}\ \bibnamefont
  {Csikor}}, \bibinfo {author} {\bibfnamefont {C.}~\bibnamefont {Motz}},
  \bibinfo {author} {\bibfnamefont {D.}~\bibnamefont {Weygand}}, \bibinfo
  {author} {\bibfnamefont {M.}~\bibnamefont {Zaiser}}, \ and\ \bibinfo {author}
  {\bibfnamefont {S.}~\bibnamefont {Zapperi}},\ }\href@noop {} {\bibfield
  {journal} {\bibinfo  {journal} {Science}\ }\textbf {\bibinfo {volume}
  {318}},\ \bibinfo {pages} {251} (\bibinfo {year} {2007})}\BibitemShut
  {NoStop}%
\bibitem [{\citenamefont {Weiss}\ and\ \citenamefont
  {Grasso}(1997)}]{weiss1997acoustic}%
  \BibitemOpen
  \bibfield  {author} {\bibinfo {author} {\bibfnamefont {J.}~\bibnamefont
  {Weiss}}\ and\ \bibinfo {author} {\bibfnamefont {J.-R.}\ \bibnamefont
  {Grasso}},\ }\href@noop {} {\bibfield  {journal} {\bibinfo  {journal} {The
  Journal of Physical Chemistry B}\ }\textbf {\bibinfo {volume} {101}},\
  \bibinfo {pages} {6113} (\bibinfo {year} {1997})}\BibitemShut {NoStop}%
\bibitem [{\citenamefont {Weiss}\ and\ \citenamefont
  {Marsan}(2003)}]{weiss2003three}%
  \BibitemOpen
  \bibfield  {author} {\bibinfo {author} {\bibfnamefont {J.}~\bibnamefont
  {Weiss}}\ and\ \bibinfo {author} {\bibfnamefont {D.}~\bibnamefont {Marsan}},\
  }\href@noop {} {\bibfield  {journal} {\bibinfo  {journal} {Science}\ }\textbf
  {\bibinfo {volume} {299}},\ \bibinfo {pages} {89} (\bibinfo {year}
  {2003})}\BibitemShut {NoStop}%
\bibitem [{\citenamefont {Weiss}\ \emph {et~al.}(2015)\citenamefont {Weiss},
  \citenamefont {Rhouma}, \citenamefont {Richeton}, \citenamefont {Dechanel},
  \citenamefont {Louchet},\ and\ \citenamefont {Truskinovsky}}]{weiss2015mild}%
  \BibitemOpen
  \bibfield  {author} {\bibinfo {author} {\bibfnamefont {J.}~\bibnamefont
  {Weiss}}, \bibinfo {author} {\bibfnamefont {W.~B.}\ \bibnamefont {Rhouma}},
  \bibinfo {author} {\bibfnamefont {T.}~\bibnamefont {Richeton}}, \bibinfo
  {author} {\bibfnamefont {S.}~\bibnamefont {Dechanel}}, \bibinfo {author}
  {\bibfnamefont {F.}~\bibnamefont {Louchet}}, \ and\ \bibinfo {author}
  {\bibfnamefont {L.}~\bibnamefont {Truskinovsky}},\ }\href@noop {} {\bibfield
  {journal} {\bibinfo  {journal} {Physical review letters}\ }\textbf {\bibinfo
  {volume} {114}},\ \bibinfo {pages} {105504} (\bibinfo {year}
  {2015})}\BibitemShut {NoStop}%
\bibitem [{\citenamefont {Isp{\'a}novity}\ \emph {et~al.}(2022)\citenamefont
  {Isp{\'a}novity}, \citenamefont {Ugi}, \citenamefont {P{\'e}terffy},
  \citenamefont {Knapek}, \citenamefont {Kal{\'a}cska}, \citenamefont
  {T{\"u}zes}, \citenamefont {Dankh{\'a}zi}, \citenamefont {M{\'a}this},
  \citenamefont {Chmel{\'\i}k},\ and\ \citenamefont
  {Groma}}]{ispanovity2022dislocation}%
  \BibitemOpen
  \bibfield  {author} {\bibinfo {author} {\bibfnamefont {P.~D.}\ \bibnamefont
  {Isp{\'a}novity}}, \bibinfo {author} {\bibfnamefont {D.}~\bibnamefont {Ugi}},
  \bibinfo {author} {\bibfnamefont {G.}~\bibnamefont {P{\'e}terffy}}, \bibinfo
  {author} {\bibfnamefont {M.}~\bibnamefont {Knapek}}, \bibinfo {author}
  {\bibfnamefont {S.}~\bibnamefont {Kal{\'a}cska}}, \bibinfo {author}
  {\bibfnamefont {D.}~\bibnamefont {T{\"u}zes}}, \bibinfo {author}
  {\bibfnamefont {Z.}~\bibnamefont {Dankh{\'a}zi}}, \bibinfo {author}
  {\bibfnamefont {K.}~\bibnamefont {M{\'a}this}}, \bibinfo {author}
  {\bibfnamefont {F.}~\bibnamefont {Chmel{\'\i}k}}, \ and\ \bibinfo {author}
  {\bibfnamefont {I.}~\bibnamefont {Groma}},\ }\href@noop {} {\bibfield
  {journal} {\bibinfo  {journal} {Nature communications}\ }\textbf {\bibinfo
  {volume} {13}},\ \bibinfo {pages} {1975} (\bibinfo {year}
  {2022})}\BibitemShut {NoStop}%
\bibitem [{\citenamefont {Spaepen}(1977)}]{spaepen1977microscopic}%
  \BibitemOpen
  \bibfield  {author} {\bibinfo {author} {\bibfnamefont {F.}~\bibnamefont
  {Spaepen}},\ }\href@noop {} {\bibfield  {journal} {\bibinfo  {journal} {Acta
  metallurgica}\ }\textbf {\bibinfo {volume} {25}},\ \bibinfo {pages} {407}
  (\bibinfo {year} {1977})}\BibitemShut {NoStop}%
\bibitem [{\citenamefont {Falk}\ and\ \citenamefont
  {Langer}(1998)}]{falk1998dynamics}%
  \BibitemOpen
  \bibfield  {author} {\bibinfo {author} {\bibfnamefont {M.~L.}\ \bibnamefont
  {Falk}}\ and\ \bibinfo {author} {\bibfnamefont {J.~S.}\ \bibnamefont
  {Langer}},\ }\href@noop {} {\bibfield  {journal} {\bibinfo  {journal}
  {Physical Review E}\ }\textbf {\bibinfo {volume} {57}},\ \bibinfo {pages}
  {7192} (\bibinfo {year} {1998})}\BibitemShut {NoStop}%
\bibitem [{\citenamefont {Kabla}\ and\ \citenamefont
  {Debr{\'e}geas}(2003)}]{kabla2003local}%
  \BibitemOpen
  \bibfield  {author} {\bibinfo {author} {\bibfnamefont {A.}~\bibnamefont
  {Kabla}}\ and\ \bibinfo {author} {\bibfnamefont {G.}~\bibnamefont
  {Debr{\'e}geas}},\ }\href@noop {} {\bibfield  {journal} {\bibinfo  {journal}
  {Physical review letters}\ }\textbf {\bibinfo {volume} {90}},\ \bibinfo
  {pages} {258303} (\bibinfo {year} {2003})}\BibitemShut {NoStop}%
\bibitem [{\citenamefont {Dennin}(2004)}]{dennin2004statistics}%
  \BibitemOpen
  \bibfield  {author} {\bibinfo {author} {\bibfnamefont {M.}~\bibnamefont
  {Dennin}},\ }\href@noop {} {\bibfield  {journal} {\bibinfo  {journal}
  {Physical Review E}\ }\textbf {\bibinfo {volume} {70}},\ \bibinfo {pages}
  {041406} (\bibinfo {year} {2004})}\BibitemShut {NoStop}%
\bibitem [{\citenamefont {Tainio}\ \emph {et~al.}(2021)\citenamefont {Tainio},
  \citenamefont {Viitanen}, \citenamefont {Mac~Intyre}, \citenamefont {Aydin},
  \citenamefont {Koivisto}, \citenamefont {Puisto},\ and\ \citenamefont
  {Alava}}]{tainio2021predicting}%
  \BibitemOpen
  \bibfield  {author} {\bibinfo {author} {\bibfnamefont {O.}~\bibnamefont
  {Tainio}}, \bibinfo {author} {\bibfnamefont {L.}~\bibnamefont {Viitanen}},
  \bibinfo {author} {\bibfnamefont {J.~R.}\ \bibnamefont {Mac~Intyre}},
  \bibinfo {author} {\bibfnamefont {M.}~\bibnamefont {Aydin}}, \bibinfo
  {author} {\bibfnamefont {J.}~\bibnamefont {Koivisto}}, \bibinfo {author}
  {\bibfnamefont {A.}~\bibnamefont {Puisto}}, \ and\ \bibinfo {author}
  {\bibfnamefont {M.}~\bibnamefont {Alava}},\ }\href@noop {} {\bibfield
  {journal} {\bibinfo  {journal} {Physical Review Materials}\ }\textbf
  {\bibinfo {volume} {5}},\ \bibinfo {pages} {075601} (\bibinfo {year}
  {2021})}\BibitemShut {NoStop}%
\bibitem [{\citenamefont {Richard}\ \emph {et~al.}(2020)\citenamefont
  {Richard}, \citenamefont {Ozawa}, \citenamefont {Patinet}, \citenamefont
  {Stanifer}, \citenamefont {Shang}, \citenamefont {Ridout}, \citenamefont
  {Xu}, \citenamefont {Zhang}, \citenamefont {Morse}, \citenamefont {Barrat}
  \emph {et~al.}}]{richard2020predicting}%
  \BibitemOpen
  \bibfield  {author} {\bibinfo {author} {\bibfnamefont {D.}~\bibnamefont
  {Richard}}, \bibinfo {author} {\bibfnamefont {M.}~\bibnamefont {Ozawa}},
  \bibinfo {author} {\bibfnamefont {S.}~\bibnamefont {Patinet}}, \bibinfo
  {author} {\bibfnamefont {E.}~\bibnamefont {Stanifer}}, \bibinfo {author}
  {\bibfnamefont {B.}~\bibnamefont {Shang}}, \bibinfo {author} {\bibfnamefont
  {S.}~\bibnamefont {Ridout}}, \bibinfo {author} {\bibfnamefont
  {B.}~\bibnamefont {Xu}}, \bibinfo {author} {\bibfnamefont {G.}~\bibnamefont
  {Zhang}}, \bibinfo {author} {\bibfnamefont {P.}~\bibnamefont {Morse}},
  \bibinfo {author} {\bibfnamefont {J.-L.}\ \bibnamefont {Barrat}},  \emph
  {et~al.},\ }\href@noop {} {\bibfield  {journal} {\bibinfo  {journal}
  {Physical Review Materials}\ }\textbf {\bibinfo {volume} {4}},\ \bibinfo
  {pages} {113609} (\bibinfo {year} {2020})}\BibitemShut {NoStop}%
\bibitem [{\citenamefont {Patinet}\ \emph {et~al.}(2016)\citenamefont
  {Patinet}, \citenamefont {Vandembroucq},\ and\ \citenamefont
  {Falk}}]{patinet2016connecting}%
  \BibitemOpen
  \bibfield  {author} {\bibinfo {author} {\bibfnamefont {S.}~\bibnamefont
  {Patinet}}, \bibinfo {author} {\bibfnamefont {D.}~\bibnamefont
  {Vandembroucq}}, \ and\ \bibinfo {author} {\bibfnamefont {M.~L.}\
  \bibnamefont {Falk}},\ }\href@noop {} {\bibfield  {journal} {\bibinfo
  {journal} {Physical review letters}\ }\textbf {\bibinfo {volume} {117}},\
  \bibinfo {pages} {045501} (\bibinfo {year} {2016})}\BibitemShut {NoStop}%
\bibitem [{\citenamefont {Berta}\ \emph {et~al.}(2023)\citenamefont {Berta},
  \citenamefont {P{\'e}terffy},\ and\ \citenamefont
  {Isp{\'a}novity}}]{berta2023dynamic}%
  \BibitemOpen
  \bibfield  {author} {\bibinfo {author} {\bibfnamefont {D.}~\bibnamefont
  {Berta}}, \bibinfo {author} {\bibfnamefont {G.}~\bibnamefont {P{\'e}terffy}},
  \ and\ \bibinfo {author} {\bibfnamefont {P.~D.}\ \bibnamefont
  {Isp{\'a}novity}},\ }\href@noop {} {\bibfield  {journal} {\bibinfo  {journal}
  {Physical Review Materials}\ }\textbf {\bibinfo {volume} {7}},\ \bibinfo
  {pages} {013604} (\bibinfo {year} {2023})}\BibitemShut {NoStop}%
\bibitem [{\citenamefont {Berta}\ \emph {et~al.}(2025)\citenamefont {Berta},
  \citenamefont {Kurunczi-Papp}, \citenamefont {Laurson},\ and\ \citenamefont
  {Ispánovity}}]{berta2025identifying}%
  \BibitemOpen
  \bibfield  {author} {\bibinfo {author} {\bibfnamefont {D.}~\bibnamefont
  {Berta}}, \bibinfo {author} {\bibfnamefont {D.}~\bibnamefont
  {Kurunczi-Papp}}, \bibinfo {author} {\bibfnamefont {L.}~\bibnamefont
  {Laurson}}, \ and\ \bibinfo {author} {\bibfnamefont {P.~D.}\ \bibnamefont
  {Ispánovity}},\ }\href@noop {} {\bibfield  {journal} {\bibinfo  {journal}
  {Acta Materialia}\ }\textbf {\bibinfo {volume} {283}},\ \bibinfo {pages}
  {120506} (\bibinfo {year} {2025})}\BibitemShut {NoStop}%
\bibitem [{\citenamefont {Barbot}\ \emph {et~al.}(2018)\citenamefont {Barbot},
  \citenamefont {Lerbinger}, \citenamefont {Hernandez-Garcia}, \citenamefont
  {Garc{\'\i}a-Garc{\'\i}a}, \citenamefont {Falk}, \citenamefont
  {Vandembroucq},\ and\ \citenamefont {Patinet}}]{barbot2018local}%
  \BibitemOpen
  \bibfield  {author} {\bibinfo {author} {\bibfnamefont {A.}~\bibnamefont
  {Barbot}}, \bibinfo {author} {\bibfnamefont {M.}~\bibnamefont {Lerbinger}},
  \bibinfo {author} {\bibfnamefont {A.}~\bibnamefont {Hernandez-Garcia}},
  \bibinfo {author} {\bibfnamefont {R.}~\bibnamefont
  {Garc{\'\i}a-Garc{\'\i}a}}, \bibinfo {author} {\bibfnamefont {M.~L.}\
  \bibnamefont {Falk}}, \bibinfo {author} {\bibfnamefont {D.}~\bibnamefont
  {Vandembroucq}}, \ and\ \bibinfo {author} {\bibfnamefont {S.}~\bibnamefont
  {Patinet}},\ }\href@noop {} {\bibfield  {journal} {\bibinfo  {journal}
  {Physical Review E}\ }\textbf {\bibinfo {volume} {97}},\ \bibinfo {pages}
  {033001} (\bibinfo {year} {2018})}\BibitemShut {NoStop}%
\bibitem [{\citenamefont {Morris}\ \emph {et~al.}(2011)\citenamefont {Morris},
  \citenamefont {Bei}, \citenamefont {Pharr},\ and\ \citenamefont
  {George}}]{morris2011size}%
  \BibitemOpen
  \bibfield  {author} {\bibinfo {author} {\bibfnamefont {J.~R.}\ \bibnamefont
  {Morris}}, \bibinfo {author} {\bibfnamefont {H.}~\bibnamefont {Bei}},
  \bibinfo {author} {\bibfnamefont {G.~M.}\ \bibnamefont {Pharr}}, \ and\
  \bibinfo {author} {\bibfnamefont {E.~P.}\ \bibnamefont {George}},\
  }\href@noop {} {\bibfield  {journal} {\bibinfo  {journal} {Physical review
  letters}\ }\textbf {\bibinfo {volume} {106}},\ \bibinfo {pages} {165502}
  (\bibinfo {year} {2011})}\BibitemShut {NoStop}%
\bibitem [{\citenamefont {Liu}\ and\ \citenamefont
  {Maa{\ss}}(2018)}]{liu2018elastic}%
  \BibitemOpen
  \bibfield  {author} {\bibinfo {author} {\bibfnamefont {C.}~\bibnamefont
  {Liu}}\ and\ \bibinfo {author} {\bibfnamefont {R.}~\bibnamefont {Maa{\ss}}},\
  }\href@noop {} {\bibfield  {journal} {\bibinfo  {journal} {Advanced
  Functional Materials}\ }\textbf {\bibinfo {volume} {28}},\ \bibinfo {pages}
  {1800388} (\bibinfo {year} {2018})}\BibitemShut {NoStop}%
\bibitem [{\citenamefont {Ovaska}\ \emph {et~al.}(2015)\citenamefont {Ovaska},
  \citenamefont {Laurson},\ and\ \citenamefont {Alava}}]{ovaska2015quenched}%
  \BibitemOpen
  \bibfield  {author} {\bibinfo {author} {\bibfnamefont {M.}~\bibnamefont
  {Ovaska}}, \bibinfo {author} {\bibfnamefont {L.}~\bibnamefont {Laurson}}, \
  and\ \bibinfo {author} {\bibfnamefont {M.~J.}\ \bibnamefont {Alava}},\
  }\href@noop {} {\bibfield  {journal} {\bibinfo  {journal} {Scientific
  reports}\ }\textbf {\bibinfo {volume} {5}},\ \bibinfo {pages} {1} (\bibinfo
  {year} {2015})}\BibitemShut {NoStop}%
\bibitem [{\citenamefont {Pan}\ \emph {et~al.}(2017)\citenamefont {Pan},
  \citenamefont {Hu},\ and\ \citenamefont {Wu}}]{pan2017generalized}%
  \BibitemOpen
  \bibfield  {author} {\bibinfo {author} {\bibfnamefont {X.}~\bibnamefont
  {Pan}}, \bibinfo {author} {\bibfnamefont {W.}~\bibnamefont {Hu}}, \ and\
  \bibinfo {author} {\bibfnamefont {C.}~\bibnamefont {Wu}},\ }\href@noop {}
  {\bibfield  {journal} {\bibinfo  {journal} {Journal of the Mechanics and
  Physics of Solids}\ }\textbf {\bibinfo {volume} {103}},\ \bibinfo {pages} {3}
  (\bibinfo {year} {2017})}\BibitemShut {NoStop}%
\bibitem [{\citenamefont {Berta}\ \emph {et~al.}(2020)\citenamefont {Berta},
  \citenamefont {Groma},\ and\ \citenamefont
  {Isp{\'a}novity}}]{berta2020efficient}%
  \BibitemOpen
  \bibfield  {author} {\bibinfo {author} {\bibfnamefont {D.}~\bibnamefont
  {Berta}}, \bibinfo {author} {\bibfnamefont {I.}~\bibnamefont {Groma}}, \ and\
  \bibinfo {author} {\bibfnamefont {P.~D.}\ \bibnamefont {Isp{\'a}novity}},\
  }\href@noop {} {\bibfield  {journal} {\bibinfo  {journal} {Modelling and
  Simulation in Materials Science and Engineering}\ }\textbf {\bibinfo {volume}
  {28}},\ \bibinfo {pages} {035014} (\bibinfo {year} {2020})}\BibitemShut
  {NoStop}%
\bibitem [{\citenamefont {Shima}\ \emph {et~al.}(2022)\citenamefont {Shima},
  \citenamefont {Sumigawa},\ and\ \citenamefont
  {Umeno}}]{shima2022nonsingular}%
  \BibitemOpen
  \bibfield  {author} {\bibinfo {author} {\bibfnamefont {H.}~\bibnamefont
  {Shima}}, \bibinfo {author} {\bibfnamefont {T.}~\bibnamefont {Sumigawa}}, \
  and\ \bibinfo {author} {\bibfnamefont {Y.}~\bibnamefont {Umeno}},\
  }\href@noop {} {\bibfield  {journal} {\bibinfo  {journal} {Materials}\
  }\textbf {\bibinfo {volume} {15}},\ \bibinfo {pages} {4929} (\bibinfo {year}
  {2022})}\BibitemShut {NoStop}%
\bibitem [{\citenamefont {P{\'e}terffy}\ and\ \citenamefont
  {Isp{\'a}novity}(2020)}]{peterffy2020efficient}%
  \BibitemOpen
  \bibfield  {author} {\bibinfo {author} {\bibfnamefont {G.}~\bibnamefont
  {P{\'e}terffy}}\ and\ \bibinfo {author} {\bibfnamefont {P.~D.}\ \bibnamefont
  {Isp{\'a}novity}},\ }\href@noop {} {\bibfield  {journal} {\bibinfo  {journal}
  {Modelling and Simulation in Materials Science and Engineering}\ }\textbf
  {\bibinfo {volume} {28}},\ \bibinfo {pages} {035013} (\bibinfo {year}
  {2020})}\BibitemShut {NoStop}%
\bibitem [{\citenamefont {Tsekenis}\ \emph {et~al.}(2011)\citenamefont
  {Tsekenis}, \citenamefont {Goldenfeld},\ and\ \citenamefont
  {Dahmen}}]{tsekenis2011dislocations}%
  \BibitemOpen
  \bibfield  {author} {\bibinfo {author} {\bibfnamefont {G.}~\bibnamefont
  {Tsekenis}}, \bibinfo {author} {\bibfnamefont {N.}~\bibnamefont
  {Goldenfeld}}, \ and\ \bibinfo {author} {\bibfnamefont {K.~A.}\ \bibnamefont
  {Dahmen}},\ }\href@noop {} {\bibfield  {journal} {\bibinfo  {journal}
  {Physical review letters}\ }\textbf {\bibinfo {volume} {106}},\ \bibinfo
  {pages} {105501} (\bibinfo {year} {2011})}\BibitemShut {NoStop}%
\bibitem [{\citenamefont {Szab{\'o}}\ \emph {et~al.}(2015)\citenamefont
  {Szab{\'o}}, \citenamefont {Isp{\'a}novity},\ and\ \citenamefont
  {Groma}}]{szabo2015plastic}%
  \BibitemOpen
  \bibfield  {author} {\bibinfo {author} {\bibfnamefont {P.}~\bibnamefont
  {Szab{\'o}}}, \bibinfo {author} {\bibfnamefont {P.~D.}\ \bibnamefont
  {Isp{\'a}novity}}, \ and\ \bibinfo {author} {\bibfnamefont {I.}~\bibnamefont
  {Groma}},\ }\href@noop {} {\bibfield  {journal} {\bibinfo  {journal}
  {Physical Review B}\ }\textbf {\bibinfo {volume} {91}},\ \bibinfo {pages}
  {054106} (\bibinfo {year} {2015})}\BibitemShut {NoStop}%
\bibitem [{\citenamefont {Zaiser}\ \emph {et~al.}(2001)\citenamefont {Zaiser},
  \citenamefont {Miguel},\ and\ \citenamefont {Groma}}]{zaiser2001statistical}%
  \BibitemOpen
  \bibfield  {author} {\bibinfo {author} {\bibfnamefont {M.}~\bibnamefont
  {Zaiser}}, \bibinfo {author} {\bibfnamefont {M.-C.}\ \bibnamefont {Miguel}},
  \ and\ \bibinfo {author} {\bibfnamefont {I.}~\bibnamefont {Groma}},\
  }\href@noop {} {\bibfield  {journal} {\bibinfo  {journal} {Physical Review
  B}\ }\textbf {\bibinfo {volume} {64}},\ \bibinfo {pages} {224102} (\bibinfo
  {year} {2001})}\BibitemShut {NoStop}%
\bibitem [{\citenamefont {Isp{\'a}novity}\ \emph {et~al.}(2008)\citenamefont
  {Isp{\'a}novity}, \citenamefont {Groma},\ and\ \citenamefont
  {Gy{\"o}rgyi}}]{ispanovity2008evolution}%
  \BibitemOpen
  \bibfield  {author} {\bibinfo {author} {\bibfnamefont {P.~D.}\ \bibnamefont
  {Isp{\'a}novity}}, \bibinfo {author} {\bibfnamefont {I.}~\bibnamefont
  {Groma}}, \ and\ \bibinfo {author} {\bibfnamefont {G.}~\bibnamefont
  {Gy{\"o}rgyi}},\ }\href@noop {} {\bibfield  {journal} {\bibinfo  {journal}
  {Physical Review B—Condensed Matter and Materials Physics}\ }\textbf
  {\bibinfo {volume} {78}},\ \bibinfo {pages} {024119} (\bibinfo {year}
  {2008})}\BibitemShut {NoStop}%
\bibitem [{\citenamefont {Groma}\ \emph {et~al.}(2006)\citenamefont {Groma},
  \citenamefont {Gy{\"o}rgyi},\ and\ \citenamefont {Kocsis}}]{groma2006debye}%
  \BibitemOpen
  \bibfield  {author} {\bibinfo {author} {\bibfnamefont {I.}~\bibnamefont
  {Groma}}, \bibinfo {author} {\bibfnamefont {G.}~\bibnamefont {Gy{\"o}rgyi}},
  \ and\ \bibinfo {author} {\bibfnamefont {B.}~\bibnamefont {Kocsis}},\
  }\href@noop {} {\bibfield  {journal} {\bibinfo  {journal} {Physical review
  letters}\ }\textbf {\bibinfo {volume} {96}},\ \bibinfo {pages} {165503}
  (\bibinfo {year} {2006})}\BibitemShut {NoStop}%
\bibitem [{\citenamefont {Derlet}\ and\ \citenamefont
  {Maa{\ss}}(2016)}]{derlet2016critical}%
  \BibitemOpen
  \bibfield  {author} {\bibinfo {author} {\bibfnamefont {P.}~\bibnamefont
  {Derlet}}\ and\ \bibinfo {author} {\bibfnamefont {R.}~\bibnamefont
  {Maa{\ss}}},\ }\href@noop {} {\bibfield  {journal} {\bibinfo  {journal}
  {Physical Review E}\ }\textbf {\bibinfo {volume} {94}},\ \bibinfo {pages}
  {033001} (\bibinfo {year} {2016})}\BibitemShut {NoStop}%
\bibitem [{\citenamefont {Isp{\'a}novity}\ \emph {et~al.}(2014)\citenamefont
  {Isp{\'a}novity}, \citenamefont {Laurson}, \citenamefont {Zaiser},
  \citenamefont {Groma}, \citenamefont {Zapperi},\ and\ \citenamefont
  {Alava}}]{ispanovity2014avalanches}%
  \BibitemOpen
  \bibfield  {author} {\bibinfo {author} {\bibfnamefont {P.~D.}\ \bibnamefont
  {Isp{\'a}novity}}, \bibinfo {author} {\bibfnamefont {L.}~\bibnamefont
  {Laurson}}, \bibinfo {author} {\bibfnamefont {M.}~\bibnamefont {Zaiser}},
  \bibinfo {author} {\bibfnamefont {I.}~\bibnamefont {Groma}}, \bibinfo
  {author} {\bibfnamefont {S.}~\bibnamefont {Zapperi}}, \ and\ \bibinfo
  {author} {\bibfnamefont {M.~J.}\ \bibnamefont {Alava}},\ }\href@noop {}
  {\bibfield  {journal} {\bibinfo  {journal} {Physical review letters}\
  }\textbf {\bibinfo {volume} {112}},\ \bibinfo {pages} {235501} (\bibinfo
  {year} {2014})}\BibitemShut {NoStop}%
\bibitem [{\citenamefont {Weibull}(1939)}]{weibull1939statistical}%
  \BibitemOpen
  \bibfield  {author} {\bibinfo {author} {\bibfnamefont {W.}~\bibnamefont
  {Weibull}},\ }\href@noop {} {\bibfield  {journal} {\bibinfo  {journal}
  {IVB-Handl.}\ } (\bibinfo {year} {1939})}\BibitemShut {NoStop}%
\bibitem [{\citenamefont {Weibull}(1951)}]{weibull1951wide}%
  \BibitemOpen
  \bibfield  {author} {\bibinfo {author} {\bibfnamefont {W.}~\bibnamefont
  {Weibull}},\ }\href@noop {} {\bibfield  {journal} {\bibinfo  {journal}
  {Journal of applied mechanics}\ }\textbf {\bibinfo {volume} {103}},\ \bibinfo
  {pages} {293} (\bibinfo {year} {1951})}\BibitemShut {NoStop}%
\bibitem [{\citenamefont {Derlet}\ and\ \citenamefont
  {Maass}(2015)}]{derlet2015probabilistic}%
  \BibitemOpen
  \bibfield  {author} {\bibinfo {author} {\bibfnamefont {P.~M.}\ \bibnamefont
  {Derlet}}\ and\ \bibinfo {author} {\bibfnamefont {R.}~\bibnamefont {Maass}},\
  }\href@noop {} {\bibfield  {journal} {\bibinfo  {journal} {Philosophical
  Magazine}\ }\textbf {\bibinfo {volume} {95}},\ \bibinfo {pages} {1829}
  (\bibinfo {year} {2015})}\BibitemShut {NoStop}%
\bibitem [{\citenamefont {Derlet}\ and\ \citenamefont
  {Maa{\ss}}(2015)}]{derlet2015universal}%
  \BibitemOpen
  \bibfield  {author} {\bibinfo {author} {\bibfnamefont {P.}~\bibnamefont
  {Derlet}}\ and\ \bibinfo {author} {\bibfnamefont {R.}~\bibnamefont
  {Maa{\ss}}},\ }\href@noop {} {\bibfield  {journal} {\bibinfo  {journal}
  {Scripta Materialia}\ }\textbf {\bibinfo {volume} {109}},\ \bibinfo {pages}
  {19} (\bibinfo {year} {2015})}\BibitemShut {NoStop}%
\bibitem [{\citenamefont {Isp{\'a}novity}\ \emph {et~al.}(2017)\citenamefont
  {Isp{\'a}novity}, \citenamefont {T{\"u}zes}, \citenamefont {Szab{\'o}},
  \citenamefont {Zaiser},\ and\ \citenamefont {Groma}}]{ispanovity2017role}%
  \BibitemOpen
  \bibfield  {author} {\bibinfo {author} {\bibfnamefont {P.~D.}\ \bibnamefont
  {Isp{\'a}novity}}, \bibinfo {author} {\bibfnamefont {D.}~\bibnamefont
  {T{\"u}zes}}, \bibinfo {author} {\bibfnamefont {P.}~\bibnamefont
  {Szab{\'o}}}, \bibinfo {author} {\bibfnamefont {M.}~\bibnamefont {Zaiser}}, \
  and\ \bibinfo {author} {\bibfnamefont {I.}~\bibnamefont {Groma}},\
  }\href@noop {} {\bibfield  {journal} {\bibinfo  {journal} {Physical Review
  B}\ }\textbf {\bibinfo {volume} {95}},\ \bibinfo {pages} {054108} (\bibinfo
  {year} {2017})}\BibitemShut {NoStop}%
\bibitem [{\citenamefont {Salmenjoki}\ \emph {et~al.}(2018)\citenamefont
  {Salmenjoki}, \citenamefont {Alava},\ and\ \citenamefont
  {Laurson}}]{salmenjoki2018machine}%
  \BibitemOpen
  \bibfield  {author} {\bibinfo {author} {\bibfnamefont {H.}~\bibnamefont
  {Salmenjoki}}, \bibinfo {author} {\bibfnamefont {M.~J.}\ \bibnamefont
  {Alava}}, \ and\ \bibinfo {author} {\bibfnamefont {L.}~\bibnamefont
  {Laurson}},\ }\href@noop {} {\bibfield  {journal} {\bibinfo  {journal}
  {Nature communications}\ }\textbf {\bibinfo {volume} {9}},\ \bibinfo {pages}
  {5307} (\bibinfo {year} {2018})}\BibitemShut {NoStop}%
\bibitem [{\citenamefont {Eshelby}(1957)}]{eshelby1957determination}%
  \BibitemOpen
  \bibfield  {author} {\bibinfo {author} {\bibfnamefont {J.~D.}\ \bibnamefont
  {Eshelby}},\ }\href@noop {} {\bibfield  {journal} {\bibinfo  {journal}
  {Proceedings of the royal society of London. Series A. Mathematical and
  physical sciences}\ }\textbf {\bibinfo {volume} {241}},\ \bibinfo {pages}
  {376} (\bibinfo {year} {1957})}\BibitemShut {NoStop}%
\end{thebibliography}%

\end{document}